\documentclass[12pt]{article}
\usepackage[T1]{fontenc}
\usepackage[utf8]{inputenc}
\usepackage{lmodern}
\usepackage[margin=1in]{geometry}
\usepackage{amsmath,amssymb,amsfonts,amsthm,mathtools}
\usepackage{booktabs,array,enumitem,etoolbox,microtype}
\usepackage[round,authoryear]{natbib}
\usepackage{xcolor}
\usepackage[colorlinks=true,linkcolor=blue!55!black,citecolor=blue!55!black,urlcolor=blue!55!black]{hyperref}
\usepackage{xurl}
\usepackage[nameinlink,noabbrev,capitalise]{cleveref}
\hypersetup{pdftitle={Audit Silence and the Capacity Trap},pdfauthor={Georgy Lukyanov},bookmarksdepth=3}
\theoremstyle{plain}
\newtheorem{theorem}{Theorem}[section]
\newtheorem{proposition}[theorem]{Proposition}
\newtheorem{lemma}[theorem]{Lemma}
\newtheorem{corollary}[theorem]{Corollary}
\newtheorem{assumption}[theorem]{Assumption}
\theoremstyle{definition}
\newtheorem{definition}[theorem]{Definition}
\theoremstyle{remark}
\newtheorem{remark}[theorem]{Remark}
\newtheorem{example}[theorem]{Example}
\newcommand{\bmhead}[1]{\subsection*{#1}}
\BeforeBeginEnvironment{proof}{\par\smallskip}
\newcommand{\E}{\mathbb{E}}
\newcommand{\Prob}{\mathbb{P}}
\newcommand{\Var}{\operatorname{Var}}
\newcommand{\Cov}{\operatorname{Cov}}

\newcommand{\al}{\underline{\alpha}}
\newcommand{\ah}{\overline{\alpha}}
\newcommand{\C}{\mathsf{C}}
\newcommand{\V}{\mathsf{V}}
\newcommand{\Lo}{\mathsf{L}}
\newcommand{\Hi}{\mathsf{H}}
\newcommand{\No}{\mathsf{N}}
\newcommand{\Ac}{\mathsf{A0}}
\newcommand{\Det}{\mathsf{D}}
\newcommand{\Tset}{\mathcal{T}}

\begin{document}
\begin{titlepage}
\centering
{\LARGE\bfseries Audit Silence and the Capacity Trap\footnote{I thank Johannes H\"orner and Allen Vong for helpful comments. Financial support from the French National Research Agency (ANR) under grant ANR-17-EURE-0010 (Investissements d'Avenir program) is gratefully acknowledged. All errors are my own.}\par}
\vspace{1.1em}
{\large Georgy Lukyanov\par}
Toulouse School of Economics, Toulouse, France\par
\href{mailto:georgy.lukyanov@tse-fr.eu}{georgy.lukyanov@tse-fr.eu}\par
\href{https://orcid.org/0009-0005-1672-610X}{ORCID: 0009-0005-1672-610X}\par
\vspace{0.7em}
September 2026\par
\vspace{0.7em}
\begin{abstract}
This paper studies how learning about an inspector's capacity affects compliance in a repeated relationship. A functioning inspector chooses effort, but the implementation of an audit remains uncertain. A constrained inspector audits at a persistently lower rate. The firm may be committed to compliance or may choose to violate; audit occurrence is observed, detection is imperfect, and a finding ends the relationship. When the constrained audit rate lies below every functioning effort level, each missed audit makes constrained capacity more likely. Under explicit payoff and belief conditions, a sufficiently long uninterrupted silence run leads to violation and maximum functioning effort in every sequential equilibrium. Further silence lowers expected auditing despite that effort. The result holds at every fixed discount factor below one, although its sufficient silence bound can become very large with patience. Entry is conditional on the specified history, and clean audits can take beliefs outside the sufficient region. We also compare technologies with the same initial maximum-effort mean audit rate. Raising the capacity floor sustains higher audit ceilings along specified silence paths and, under additional incentive and social-cost conditions, lowers expected loss over a common finite horizon. Finally, constrained inspectors become more prevalent among surviving strategic firms, while the strategic/constrained pair's share of all survivors can fall.
\end{abstract}
\begin{minipage}{\textwidth}
\par\medskip
\end{minipage}
\end{titlepage}
\section{Introduction}\label{sec:introduction}

A firm that has not been audited for some time has a reason to reconsider the enforcement it is likely to face. The absence of an audit may suggest that the inspector is unwilling to investigate. It may also suggest that the inspector has limited capacity to carry out an investigation, even when doing so would be worthwhile. This distinction matters for the firm's incentives. An inspector who is willing to act may nevertheless provide little deterrence if the firm regards successful implementation as unlikely. The present paper studies how this uncertainty about capacity affects compliance and auditing in a repeated relationship.

We consider a firm and an inspector, both of whom may have a persistent private type. The firm is either committed to compliance or chooses whether to violate a regulation. The inspector is either constrained to implement audits at a low fixed rate or can choose between two levels of effort. For the latter type, which we call the functioning inspector, higher effort raises the probability that an audit takes place. Implementation remains uncertain under either effort choice. The players observe whether an audit occurs and, if it does, whether a violation is detected. Detection is imperfect, and a detected violation ends the relationship.

The distinction between an absent audit and a clean audit is essential. When there is no audit, the firm has provided no direct evidence about its conduct. The observation does, however, convey information about the inspector. A clean audit provides evidence about both parties: it shows that the inspector managed to implement an audit, and its lack of a finding can make committed compliance more plausible. Thus a record of audit occurrence carries information that would be lost in a record containing only enforcement findings. Our main analysis assumes that the players observe this fuller record.

The central result is that a sequence of missed audits can lead to violation by every strategic firm even though every functioning inspector exerts maximum effort. We refer to this combination of behavior and weak expected enforcement as a capacity trap. The reason is as follows. Suppose that the constrained inspector's audit rate lies below the rate attainable by a functioning inspector even under low effort. Each missed audit then makes constrained capacity more likely, whatever effort the functioning type would have chosen. After sufficiently many such observations, the firm expects a low audit probability even if the functioning inspector is trying as hard as possible. Under the conditions of the theorem, violation becomes strictly profitable despite the risk of losing future payoffs. The resulting violation rate in turn makes high functioning effort strictly optimal.

There is no decline in either type's physical capacity along this history. The constrained inspector continues to audit at its fixed rate, and the functioning inspector uses its highest available rate once the sufficient conditions hold. What declines is the audit probability conditional on the information shared by the players: more weight is placed on the constrained type. This explains why greater willingness to audit need not restore deterrence. Further missed audits preserve the two actions, lower the expected detection rate, and increase the probability of an undetected violation.

The argument separates the information conveyed by silence from the incentives to act on that information. Capacity separation gives the direction of learning. To determine behavior, we also need to control the continuation payoffs that may make the firm reluctant to violate or the inspector reluctant to audit. We first use simple bounds on those payoffs. We then give a second argument that applies at every fixed discount factor below one under payoff restrictions that do not themselves depend on discounting. A positive audit floor limits the firm's continuation value. Sufficiently pessimistic capacity beliefs consequently determine its conduct over a finite set of future histories, including histories with clean audits. Over that interval, the inspector's problem can be compared with a decision problem in which firm conduct is fixed. This comparison establishes the incentive for high effort.

Both results apply to every sequential equilibrium, allowing strategies to depend on the players' private action histories. Their conclusions nevertheless concern specified histories. The theorem supplies a finite silence length sufficient for entry from a history satisfying the compliance-belief restriction; it does not show that entry occurs with probability one. Moreover, the bound for patient players can be very large. We report its size in an example because existence of a finite bound and a useful quantitative prediction about entry are different conclusions.

A clean audit can interrupt the process. It improves the firm's reputation for compliance and the inspector's reputation for functioning capacity, so both belief changes move against the sufficient conditions for the trap. We characterize when one clean audit leaves those conditions intact and when it causes departure. If only the capacity belief leaves the region, subsequent silence can restore membership. If the compliance belief exceeds the chosen region's cutoff, that same region cannot be reached again along an active history, since the compliance belief never falls. These are statements about a specified sufficient region: behavior outside it remains to be determined by the continuation equilibrium. A finite sequence of clean audits can take beliefs outside any of the fixed regions used in the argument, so the regions are not absorbing.

We then compare enforcement technologies. Holding the initial beliefs and maximum-effort mean audit rate fixed, a higher capacity floor must be accompanied by a lower peak rate. This reduction in capacity dispersion sustains a higher audit ceiling after each specified high-effort silence history. Its implications for entry bounds depend on which sufficient argument is used: the simple bound becomes longer, whereas the patient-player bound can become shorter. Neither comparison orders actual equilibrium entry times. A welfare comparison requires an additional step. We identify initial beliefs under which behavior is determined after every active history over the same finite horizon under both technologies. When the social benefit of detection is sufficiently large relative to audit and termination costs, capacity compression lowers expected loss over that horizon.

Finally, we consider what an observer can infer from the duration of a relationship. Conditional on the complete audit record, the two private types remain independent. Observing survival alone pools clean audits with missed audits and creates an association between them. Over a horizon on which the trap behavior is justified, constrained inspectors become more prevalent among surviving strategic firms. The strategic-firm/constrained-inspector pair need not, however, become a larger share of the whole surviving population. Committed firms are never detected and also accumulate among survivors. We give an exact condition for the harmful pair's population share to increase and an example in which it falls even though the relationship starts inside the sufficient trap region.

\subsection*{Related literature}
\enlargethispage{5pt}

The paper is closely related to work on learning about the effectiveness of monitoring. In \citet{CoronaRandhawa2010}, an auditor can experience a reputational slippery slope. \citet{MarinovicSzydlowski2022} study uncertain monitoring ability supported by costly investment: unpunished shirking reveals ineffectiveness and discourages further investment. In \citet{CheynelCianciarusoZhou2024}, nondetection lowers perceived monitor effectiveness and leads to escalating fraud; greater detection intensity can also accelerate learning about an ineffective monitor. These papers already establish that an absence of findings can weaken deterrence through beliefs about the monitor. An earlier working-paper extension of \citet{CheynelCianciarusoZhou2024} also includes honest and dishonest managers and endogenous investment in monitoring capacity; nondetection eventually makes continued investment unattractive.\footnote{See Section 5.3 and Proposition 9 of the \href{https://accounting.wharton.upenn.edu/wp-content/uploads/2023/04/CCZ202303_Fraud_Power_Laws-Frank-Z.pdf}{working-paper version hosted by Wharton}. This comparison refers to that version.} The presence of behavioral firm types and a discretionary monitoring margin is therefore not by itself a distinction from that work.

Our analysis focuses on the relation between the functioning inspector's effort and the enforcement expected by the firm. Persistent uncertainty about implementation capacity allows low expected enforcement to coexist with maximum functioning effort. The distinction between audit occurrence and audit findings determines the updating process, while the continuation-value argument identifies histories that force this behavior in every sequential equilibrium. The clean-audit analysis explains why the same behavior need not persist after another kind of nonterminal observation. The substantive prediction concerns the persistence of high effort along silence histories. A missed audit leaves the inspector's belief about committed compliance unchanged, so it does not generate the same reason to withdraw effort as growing confidence in an honest firm would. Once the sufficient region is reached, each further missed audit preserves maximum functioning effort even as expected enforcement falls. Clean audits can end this guarantee because they also improve the firm's reputation for compliance. The continuation-value argument establishes these conclusions for every sequential equilibrium at the specified histories, for each fixed discount factor below one under the stated restrictions. Bayesian learning from silence itself is not new.

A second group of papers studies dynamic enforcement and the design of monitoring. \citet{DilmeGarrett2019} examine residual deterrence after public enforcement successes, and \citet{VarasMarinovicSkrzypacz2020} characterize optimal dynamic inspection when monitoring affects incentives and information. \citet{MarinovicSzydlowski2023} analyze disclosure about a reputation-concerned monitor, while \citet{Tan2023} studies information about monitoring capability. \citet{ChenEtAl2026} consider how firms test a sequentially rational regulator and how transparency affects what firms learn from one another. The capacity comparison in the present paper changes the implementation technology while holding the specified initial beliefs and mean audit rate fixed. Interpreting such a change as a policy requires a resource specification; our comparison does not solve an optimal disclosure or capacity-allocation problem.

The model also relates to costly verification and enforcement. Classical contributions study the level and allocation of monitoring \citep{Townsend1979,MookherjeePng1989,MookherjeePng1992,MookherjeePng1994}, and \citet{Krawczyk2009} considers costly auditing in repeated insurance-fraud interaction. Here the occurrence of an audit additionally informs the firm about the monitoring technology. The distinction between detecting violations and having the capacity to enforce, emphasized in the tax-compliance study of \citet{Okunogbe2021}, helps motivate this separation of margins. Applying our model to a particular institution would also require its audit record and termination rules to fit the information and payoff assumptions.

Finally, the presence of two privately informed players connects the analysis to reputation and Bayesian repeated games \citep{KrepsWilson1982,FudenbergLevine1989,SalomonForges2015,MailathSamuelson2006}. \citet{Ozdogan2014} studies reputation sustainability with two long-lived privately informed players, while \citet{CrippsMailathSamuelson2004} examine the impossibility of permanently sustaining certain noncredible reputations under imperfect monitoring. Our result concerns a finite public history with positive probability in each discounted game. It does not establish a permanent false reputation or characterize long-run equilibrium play. The gain from the approach is that the specified behavior follows in every equilibrium once the sufficient conditions are met. Its cost is that play outside those conditions is left open.\footnote{In particular, a statement valid at every fixed discount factor below one need not provide a bound that is uniform as the discount factor approaches one. The distinction is visible in the numerical illustration of Theorem~\ref{thm:patient}.}

Section~\ref{sec:model} presents the model. Section~\ref{sec:beliefs} derives beliefs and incentives and defines a capacity-trap history. Section~\ref{sec:global} proves the two sufficient entry results and examines clean-audit departures. Sections~\ref{sec:design} and~\ref{sec:survivors} study capacity comparisons, welfare, and survivor selection. Section~\ref{sec:implications} discusses interpretation and scope. The main text retains the filtering and central entry proofs; Appendix C gives the static and two-period benchmarks, and Appendix F contains the supporting control-problem proofs.

\section{Model}\label{sec:model}

Consider a firm whose compliance is monitored by an inspector. In each period the firm decides whether to comply with a regulation, while the inspector can try to increase the probability of an audit. The distinction between this choice and the audit's implementation is the source of uncertainty in the model. The inspector knows whether it has functioning capacity; the firm learns about that capacity from the audits that occur, or fail to occur, during the relationship.

\subsection{Players, actions, and capacity}

Time is discrete, \(t=0,1,\ldots\). Both players discount future payoffs by \(\delta\in(0,1)\). At date zero, Nature draws their types independently. Each player observes its own type but not the other's. The prior distributions and all remaining primitives are common knowledge.

With probability \(\lambda_0\in(0,1)\), the firm is committed to compliance: it chooses \(\C\) in every period. Otherwise it is strategic and chooses between compliance, \(\C\), and violation, \(\V\). The committed type represents a firm whose conduct does not respond to the enforcement incentives considered here. For example, compliance may be secured by internal requirements outside the inspector's control. This interpretation is a maintained simplification. The strategic firm's conduct can affect the inspector's belief that it faces such a committed firm.

With probability \(\mu_0\in(0,1)\), the inspector is capacity constrained. It implements an audit with probability \(q\in(0,1)\) each period and has no discretionary effort choice. Otherwise it is functioning and chooses low effort, \(\Lo\), or high effort, \(\Hi\). Conditional on that choice, the audit probability is
\[
\theta(\Lo)=\al,\qquad \theta(\Hi)=\ah,
\qquad 0<\al<\ah<1.
\]
The random implementation draws are independent over time conditional on type and current actions. Capacity is persistent because the inspector's type does not change. Audit occurrence is stochastic even conditional on a fixed type and effort.

One interpretation is that the functioning inspector can initiate a case and pursue the procedures needed to complete it. Access to information or administrative constraints can still prevent an audit from taking place. The constrained type represents a persistent limitation for which this discretionary margin is unavailable. Thus the two types differ in the opportunities available to them, and the functioning type also makes an effort choice. The type remains fixed after a missed audit; what can change is the firm's assessment of which type it faces.

It is useful to distinguish three objects. Conditional on the functioning inspector's realized effort, the implementation probability is either \(\al\) or \(\ah\). If it chooses high effort with probability \(\rho\), the probability before that randomization is resolved is
\begin{equation}\label{eq:x}
x=\al+(\ah-\al)\rho.
\end{equation}
The probability \(x\) averages over the functioning inspector's current effort choice. The public observation comes one step later: an audit either occurs or does not. This leaves scope for learning even when effort is known to be high, since both types can generate either audit-incidence outcome. The cost specified below is paid when an audit is implemented. There is no separate direct cost of choosing high effort, although that choice raises the expected audit cost by making implementation more likely.

\begin{assumption}\label{ass:separation}
The implementation probabilities satisfy
\[
0<q<\al<\ah<1.
\]
\end{assumption}

We maintain this ordering for the infinite-horizon entry theorem. It says that even low functioning effort produces audits more frequently than the constrained technology. Consequently, observing no audit is evidence in favor of constrained capacity regardless of functioning effort. This is a substantive restriction on the technologies. If \(q=\al\), silence is uninformative when functioning effort is low; if \(q>\al\), it then favors the functioning type. The ordering is not needed for every filtering calculation, and its relaxation requires additional analysis of equilibrium effort.

\subsection{Timing, public observations, and payoffs}

A period is \emph{active} if no violation has yet been detected. In each active period the sequence is:
\begin{enumerate}[label=(\roman*),leftmargin=2.2em]
\item The strategic firm chooses \(\C\) or \(\V\), and the functioning inspector simultaneously chooses \(\Lo\) or \(\Hi\). Neither observes the other's current action.
\item An audit is implemented according to the inspector's type and effort. Both players observe whether it occurs.
\item If the firm violated and an audit occurs, the violation is detected with probability \(d\in(0,1)\). A compliant firm never produces a finding. Any finding is publicly observed.
\end{enumerate}

The public signal is \(\No\) when no audit occurs, \(\Ac\) when an audit occurs with no finding, and \(\Det\) when a violation is detected. We call \(\Ac\) a clean audit, although it need not establish compliance: an actual violation can go undetected. Audit silence refers only to \(\No\).

The distinction between these two nonterminal signals requires a record that identifies audit occurrence, including audits without findings. The regulated firm and inspector are assumed to have this record. Any outside observer to whom we apply the public-history predictions must have it as well. A researcher may instead observe only whether the relationship remains active; that is a different information set, considered in Section~\ref{sec:survivors}. A list of enforcement findings alone does not supply the information assumed in the main analysis.

A finding ends the relationship and gives both players zero subsequent payoff. This assumption can describe removal of the firm from the monitored activity, termination of a contract, or a reset whose value is included in the terminal payoff. It makes detection costly to a strategic firm partly because detection removes future opportunities to earn from violations. The model therefore applies to a relationship with this termination feature. Routine penalties followed by continuation of the same relationship would require a different specification.

Effort affects implementation but not the conditional detection probability \(d\). This separates initiating and completing an audit from the accuracy of the examination once implemented. If audit incidence were fixed and identical across inspector types, its absence would provide no capacity information. If incidence differed across types but effort instead affected detection accuracy, silence could still be informative; however, the clean-audit likelihood would generally depend jointly on firm conduct and inspector effort. The factorization used below would then require a new argument. We do not treat those alternatives as equivalent to the present model.

An implemented audit costs the inspector \(c>0\). An undetected violation gives the firm \(g>0\) and imposes loss \(\ell>0\) on the inspector. Detection gives the firm \(-f\), where \(f>0\), and gives the inspector \(r-c\), where \(r\geq0\). Table~\ref{tab:payoffs} records the resulting payoffs.

\begin{table}[t]
\centering
\caption{Outcomes and active-period payoffs}
\label{tab:payoffs}
\begin{tabular}{@{}lllcc@{}}
\toprule
Firm action & Enforcement outcome & Public signal & Firm & Inspector\\
\midrule
\(\C\) & no audit & \(\No\) & \(0\) & \(0\)\\
\(\C\) & audit & \(\Ac\) & \(0\) & \(-c\)\\
\(\V\) & no audit & \(\No\) & \(g\) & \(-\ell\)\\
\(\V\) & audit, false negative & \(\Ac\) & \(g\) & \(-\ell-c\)\\
\(\V\) & audit, detected & \(\Det\) & \(-f\) & \(r-c\)\\
\bottomrule
\end{tabular}
\end{table}

Both strategic players maximize expected unnormalized discounted sums of these payoffs, with zero continuation after detection. Multiplication of the entire objective by \(1-\delta\) would leave preferences unchanged. The zero post-detection continuation is part of our specification; a constant terminal continuation could instead be incorporated into the detection payoffs, subject to the stated sign restrictions.

The players receive no private information beyond their own types and actions. In particular, the inspector does not observe the realized loss \(\ell\) as an additional signal before choosing its next action. The loss measures the consequence of an undetected violation in the inspector's objective; it does not reveal that violation. Allowing it to reveal conduct would change both the inspector's information and the filtering argument below.\footnote{The distinction is between utility and information. A payoff can represent a consequence that the player cares about without assuming that the player observes that consequence separately from the signals specified in the game.}

\subsection{Histories, strategies, and equilibrium}
\enlargethispage{2pt}

Let \(y_t\in\{\No,\Ac,\Det\}\) denote the public signal in period \(t\). An active public history at the beginning of period \(t\) is
\[
h_t=(y_0,\ldots,y_{t-1})\in\{\No,\Ac\}^{t}.
\]
The strategic firm's information consists of \(h_t\), its type, and its own action sequence \(z^{t}=(z_0,\ldots,z_{t-1})\). The functioning inspector similarly knows its type, \(h_t\), and its effort sequence \(e^{t}=(e_0,\ldots,e_{t-1})\). A behavioral strategy assigns an action probability at each such information set. The players may condition on their entire private histories; no public-strategy or Markov restriction is imposed in the entry theorem.

An assessment consists of these behavioral strategies and beliefs over the nodes in each information set. We use sequential equilibrium: continuation strategies must maximize expected discounted payoffs at every information set, and beliefs must be limits of Bayesian beliefs induced by completely mixed behavioral strategies. The perturbations apply to the strategic players' available actions; they do not change the committed firm's prescribed conduct or the constrained technology. Consistency is understood pointwise at each finite history. Bounded discounted payoffs and finite action and signal sets permit a truncation-and-limit existence argument, given in the Online Appendix. The equilibrium set is therefore nonempty.

For a fixed assessment, define the public posteriors
\[
\lambda_t(h_t)=\Prob(\text{committed firm}\mid h_t),\qquad
\mu_t(h_t)=\Prob(\text{constrained inspector}\mid h_t).
\]
Let \(\sigma_t(h_t)\) be the probability of violation conditional on a strategic firm and \(h_t\), averaging over that firm's private histories. Define \(\rho_t(h_t)\) analogously as the probability of high effort conditional on a functioning inspector and \(h_t\). At a specified history we suppress the arguments and write
\[
a=(1-\lambda)\sigma,\qquad x=\al+(\ah-\al)\rho,
\]
\begin{equation}\label{eq:p}
p=\mu q+(1-\mu)x.
\end{equation}
The probability \(a\) includes the possibility that the firm is committed, whereas \(\sigma\) conditions on a strategic firm. Similarly, \(p\) includes both inspector types, whereas \(x\) conditions on functioning capacity. These distinctions will matter when we compare the functioning inspector's effort with the auditing expected by the firm. The averages are defined for an arbitrary assessment; they do not require players with different private histories to use the same strategy.

\section{Beliefs and incentives}\label{sec:beliefs}

\subsection{Public and private conditioning}

We first determine what each player knows about its opponent after a public history. The issue is slightly more demanding than applying Bayes' rule to two type probabilities. Each strategic player also remembers its own actions, and its future strategy may depend on them. We therefore need to know whether a private action history provides additional information about the opponent. The next lemma shows that it does not in this model. The reason is that the likelihood of each active public signal separates into one factor for the firm and one for the inspector.

\begin{lemma}\label{lem:beliefs}
Fix any behavioral strategy profile. After every active public history \(h_t\), the joint conditional distribution of the firm's type and private action history and the inspector's type and private action history is a product of its two marginals. In particular, the persistent types remain independent. At private histories having positive probability, each player's conditional distribution of its opponent's type and private history is the corresponding public marginal. In a sequential equilibrium the same conclusion holds at zero-probability private histories under the consistent limiting beliefs.

With \(a\) and \(p\) defined at \(h_t\) as above, the public updates are
\begin{align}
\lambda^{\Ac}&=\frac{\lambda}{1-da},&
\mu^{\Ac}&=\frac{\mu q}{p},\label{eq:update-a}\\
\lambda^{\No}&=\lambda,&
\mu^{\No}&=\frac{\mu(1-q)}{1-p}.\label{eq:update-n}
\end{align}
These formulas apply to the history-averaged actions of arbitrary behavioral strategies, not only to public strategies.
\end{lemma}

\begin{proof}
The claim holds initially because types are drawn independently. Suppose the two private states are independent conditional on an active public history. Conditional on those states, the players randomize independently. For firm action \(z\in\{\C,\V\}\) and inspector implementation probability \(\theta\), the likelihoods of the nonterminal signals are
\[
\Prob(\No\mid z,\theta)=1\cdot(1-\theta),\qquad
\Prob(\Ac\mid z,\theta)=\{1-d\mathbf{1}_{\{z=\V\}}\}\theta.
\]
Each likelihood factors into a term depending only on the firm and a term depending only on the inspector. Multiplying by the two independent private-state probabilities and the players' action probabilities preserves that product form. Normalizing after either public signal again gives a product distribution over the augmented private histories. Induction proves the claim for every active public history.

Integrating over private histories gives audit probability \(p\) independently of firm conduct. Conditional on an audit, no finding has probability \(1-da\). A committed firm always produces no finding, whereas the constrained inspector's audit probability is \(q\). Bayes' rule therefore gives \eqref{eq:update-a}. The absence of an audit has probability independent of firm conduct and probability \(1-q\) conditional on constrained capacity, giving \eqref{eq:update-n}.

Under every completely mixed perturbation, the same product identity holds at every private history and the opponent marginal does not depend on one's own private history. At any fixed finite history these are finite probability distributions. Passing to the perturbation limit preserves the identity and gives the claim at zero-probability private histories in a consistent assessment. Every nonterminal public history has positive probability, since both nonterminal signals have positive probability under every action and type pair.
\end{proof}

No audit changes only the capacity belief. A clean audit changes both beliefs: it favors the functioning inspector, which is more likely to implement the audit, and the committed firm, which is more likely to produce no finding. In particular, \(\lambda\) never decreases along an active history. It increases strictly after a clean audit whenever aggregate violation at the preceding history is positive.

The conditioning event is the complete sequence of public signals. For each such sequence, the posterior over the two private types is a product. An observer who learns only that the relationship is still active has less information: that observer must average over different sequences of clean and absent audits. The resulting average need not remain a product. We return to this distinction when studying selection among surviving relationships.

\begin{lemma}\label{lem:private-values-main}
In a sequential equilibrium, a strategic player's continuation value is the same at any two of its private information sets with the same active public history and own type. The conclusion includes consistent beliefs at zero-probability private histories.
\end{lemma}

\begin{proof}
By Lemma~\ref{lem:beliefs}, the distribution of the opponent's type and private history is identical at the two information sets. Own past actions are unobserved by the opponent and do not affect future payoffs or implementation probabilities. Every feasible continuation plan at one information set has a corresponding plan at the other: use the same choices after every subsequent sequence of own actions and public signals. Against the opponent's fixed strategy, corresponding plans yield the same expected continuation payoff. The two optimization problems consequently have the same value. Sequential rationality requires that value to be attained at each information set.
\end{proof}

The lemma allows us to write a continuation value for each own type and public history even when strategies use private histories. It does not require the strategies themselves to coincide: if several actions give the same optimal value, different private histories may select different actions. Nor can we replace the public history by the pair of marginal beliefs without another argument. Two public histories with the same beliefs may lead to different future equilibrium play.

\subsection{Dynamic deviation gains}

Fix an active strategic-firm information set. By Lemma~\ref{lem:beliefs}, its audit probability equals the public-history average \(p\). Let \(V_{\Ac}\) and \(V_{\No}\) denote the strategic firm's equilibrium continuation values after the indicated public signals. Lemma~\ref{lem:private-values-main} ensures that those values do not depend on whether the firm reached the successor after compliance or violation, including after a current deviation.

Write \(Q_{\C}\) for the expected discounted payoff from choosing compliance now and then continuing optimally, and \(Q_{\V}\) for the corresponding payoff from violation. Then
\[
Q_{\C}=\delta\{pV_{\Ac}+(1-p)V_{\No}\},
\]
\[
Q_{\V}=(1-p)(g+\delta V_{\No})
+p\{(1-d)(g+\delta V_{\Ac})-df\}.
\]
The gain from violation is
\begin{equation}\label{eq:firm-diff}
\Delta_F=Q_{\V}-Q_{\C}
=g-pd(g+f)-\delta pdV_{\Ac}.
\end{equation}
The first two terms in \eqref{eq:firm-diff} give the current gain from violation. The final term reflects the loss of continuation following a detected violation. Compliance allows the relationship to continue even when an audit occurs; violation may instead lead to detection and termination. Following no audit, either action gives the same continuation value, so that continuation cancels from the comparison. The firm may consequently comply even when its current expected gain from violation is positive, provided the continuation it preserves is sufficiently valuable.

For a functioning-inspector information set, let \(W_{\Ac}\) and \(W_{\No}\) be its continuation values. Its violation probability is the history average \(a\). Conditional on an implemented audit and on no audit, respectively, its expected values are
\begin{align*}
B_A&=-c+a\{dr-(1-d)\ell\}+\delta(1-da)W_{\Ac},\\
B_N&=-a\ell+\delta W_{\No}.
\end{align*}
High effort raises the probability of the first outcome relative to the second by \(\ah-\al\). Its payoff gain is therefore \((\ah-\al)\Psi\), where
\begin{equation}\label{eq:inspector-diff}
\Psi=-c+ad(r+\ell)+\delta\{(1-da)W_{\Ac}-W_{\No}\}.
\end{equation}
An additional chance of an audit gives the inspector an opportunity to detect misconduct, receive \(r\), and avoid the loss \(\ell\). This benefit is weighted by the probability \(a\) of violation and the detection probability \(d\); the audit itself costs \(c\). The continuation term adds two effects. A clean audit changes what the inspector believes about the firm, while a finding ends the relationship. The signs of these continuation effects cannot be settled by the current payoffs alone. This is why establishing high effort in the repeated game requires a bound or a comparison of continuation values.

\subsection{The deterrence benchmark and the capacity trap}

Define
\begin{equation}\label{eq:static-thresholds}
p^*=\frac{g}{d(g+f)},\qquad
a^*=\frac{c}{d(r+\ell)}.
\end{equation}

With no continuation payoff, the strategic firm prefers violation below \(p^*\), and the functioning inspector prefers high effort above \(a^*\). These thresholds identify the current-payoff tension: if even maximum functioning effort gives \(p^H(\mu)=\mu q+(1-\mu)\ah<p^*\), low expected enforcement can coexist with a strong incentive to audit. In the repeated game, however, continuation payoffs must also be taken into account. The next section establishes conditions under which they cannot reverse either action.

\begin{definition}\label{def:trap}
An active public history is a \emph{capacity-trap history} in a specified sequential equilibrium if every strategic-firm information set at that history prescribes violation, every functioning-inspector information set prescribes high effort, and the resulting audit probability \(p^H(\mu)\) is below the one-period deterrence threshold \(p^*=g/[d(g+f)]\).
\end{definition}

At a capacity-trap history, the functioning inspector chooses its highest available audit rate, but the firm still expects enforcement below the one-period deterrence threshold. The definition refers to behavior at the specified history. High effort may already have been used in the preceding period, and a different signal in the next period may change the incentives of either player. We will distinguish this behavioral definition from the sets of beliefs that are sufficient to establish it.

Appendix C gives the terminal equilibrium and an exact two-period construction. In that example, a missed audit leads to a terminal capacity trap whereas a clean audit leads to interior play. The construction illustrates a possible equilibrium transition; it is not needed for the every-equilibrium sufficient conditions below.

\section{Entry and exit after audit observations}\label{sec:global}

We now ask how a relationship can reach a capacity-trap history. There are two parts to the argument. First, we identify beliefs under which continuation incentives cannot prevent strategic violation and high functioning effort. Second, we show that a sufficiently long run of missed audits brings the capacity belief into that range, provided the compliance belief satisfies the relevant restriction. The first part concerns incentives; the second uses the information contained in silence.

We begin with a direct comparison using upper and lower bounds on continuation payoffs. This gives a simple sufficient region and a finite silence bound, but its restrictions may fail when players are patient. We then use the positive audit floor and the structure of the inspector's continuation problem to obtain a result for every fixed discount factor below one. The second argument permits greater patience at the cost of a potentially much longer silence bound. Keeping both results makes their roles explicit: one gives a transparent bound where its restrictions hold, and the other establishes the broader qualitative scope of the mechanism.

\subsection{A simple sufficient region}

The strategic firm can guarantee zero by complying forever, and no period gives it more than \(g\). Hence its continuation value satisfies
\[
0\leq V\leq\frac{g}{1-\delta}.
\]
Substitution into \eqref{eq:firm-diff} shows that violation is strictly optimal whenever
\begin{equation}\label{eq:pi}
p<\pi_\delta\equiv
\frac{g(1-\delta)}{d\{g+(1-\delta)f\}}.
\end{equation}
To obtain this condition, we allow the continuation following a clean audit to be as valuable to the firm as the upper bound permits. This is the case most favorable to current compliance. If violation is still strictly profitable, a different continuation equilibrium cannot overturn that incentive. The threshold is therefore sufficient for violation at every information set with the specified audit probability.

For the functioning inspector define
\begin{equation}\label{eq:kappa-R}
\kappa_L=\al c+\max\{0,\ell-\al d(r+\ell)\},
\qquad R=\max\{r-c,0\}.
\end{equation}
Choosing low effort forever guarantees a continuation payoff of at least \(-\kappa_L/(1-\delta)\): at any violation probability \(a\), its expected current payoff under low effort is
\[
-\al c+a\{\al d(r+\ell)-\ell\}\geq-\kappa_L.
\]
Conversely, all nonterminal inspector payoffs are nonpositive. Its only possibly positive payoff is \(r-c\) at detection, and this can occur only once. Thus
\[
-\frac{\kappa_L}{1-\delta}\leq W\leq R.
\]
Applying these bounds to \eqref{eq:inspector-diff} gives
\[
\Psi\geq-c+ad(r+\ell)
-\delta\left\{\frac{\kappa_L}{1-\delta}+R\right\}.
\]
In particular, high effort is strictly optimal whenever
\begin{equation}\label{eq:beta}
a>\beta_\delta\equiv
\frac{c+\delta\{\kappa_L/(1-\delta)+R\}}{d(r+\ell)}.
\end{equation}
The inspector comparison uses the same logic. It makes the continuation after an audit as unfavorable, and the continuation after no audit as favorable, as the bounds allow. Above \(\beta_\delta\), the current benefit of detecting violations is large enough to justify high effort even under that comparison. Below this threshold the comparison is inconclusive; the actual continuation values may still support high effort. The same qualification applies to the firm's sufficient threshold.

We impose
\begin{equation}\label{eq:global-primitives}
q<\pi_\delta<\ah,\qquad \beta_\delta<1,
\end{equation}
and define
\begin{equation}\label{eq:mudagger}
\mu^\dagger=\frac{\ah-\pi_\delta}{\ah-q}.
\end{equation}
The largest audit probability feasible at belief \(\mu\) is
\[
p^H(\mu)=\mu q+(1-\mu)\ah.
\]
It is below \(\pi_\delta\) precisely when \(\mu>\mu^\dagger\).

Define the \emph{sufficient dominance region} by
\begin{equation}\label{eq:Tset}
\Tset=\{(\lambda,\mu)\in(0,1)^2:
\lambda<1-\beta_\delta,\ \mu>\mu^\dagger\}.
\end{equation}
The two inequalities in \eqref{eq:Tset} determine behavior in sequence. The restriction on \(\mu\) puts even the largest feasible audit rate below the firm's sufficient threshold. Every strategic firm therefore violates. This fixes the aggregate violation probability at \(1-\lambda\), and the restriction on \(\lambda\) puts that probability above the inspector's sufficient threshold. High functioning effort is then strictly optimal. Since \(\pi_\delta<p^*\) for \(\delta>0\), the resulting audit rate is also below the one-period deterrence threshold. Every history in \(\Tset\) is consequently a capacity-trap history in every sequential equilibrium, although such behavior may also occur outside the region.

\subsection{The common learning argument}

\begin{theorem}\label{thm:global}
Maintain Assumption~\ref{ass:separation} and \eqref{eq:global-primitives}. Fix a sequential equilibrium and an active public history with beliefs \((\lambda,\mu)\in(0,1)^2\) satisfying \(\lambda<1-\beta_\delta\). Set
\[
O=\frac{\mu}{1-\mu},\qquad
O^\dagger=\frac{\mu^\dagger}{1-\mu^\dagger},\qquad
F=\frac{1-q}{1-\al}>1,
\]
\begin{equation}\label{eq:K}
K=\min\{k\in\mathbb Z_{\geq0}:OF^k>O^\dagger\}.
\end{equation}
If the next \(K\) observations are all \(\No\), their successor belongs to \(\Tset\). At every strategic information set at that successor, violation and high functioning effort, respectively, are strict best responses. Every further observation of \(\No\) preserves membership in \(\Tset\). Conditional on the starting public history, the specified run has probability at least \(\mu(1-q)^K>0\).
\end{theorem}

\begin{proof}
At any active public history, let \(x_t\) be the functioning inspector's audit probability averaged over its private histories. It belongs to \([\al,\ah]\), irrespective of the strategies generating those histories. By Lemma~\ref{lem:beliefs}, a missed audit updates the public capacity odds according to
\[
O_{t+1}=O_t\frac{1-q}{1-x_t}\geq O_tF,
\]
and leaves \(\lambda\) unchanged. Repeating the inequality along the specified run yields \(O_K\geq OF^K>O^\dagger\). Consequently \(\mu_K>\mu^\dagger\) and \(p^H(\mu_K)<\pi_\delta\).

The private-history conclusion of Lemma~\ref{lem:beliefs} gives the same opponent-type posterior at every strategic-firm information set sharing this public history. Its audit probability cannot exceed \(p^H(\mu_K)\). The firm bound therefore makes violation strictly optimal at every such information set. Each functioning-inspector information set then faces aggregate violation \(1-\lambda>\beta_\delta\), so the inspector bound makes high effort strictly optimal. This proves membership and the behavioral claim without restricting private-history dependence before entry.

Inside the region, another missed audit multiplies the capacity odds by \((1-q)/(1-\ah)>1\) while leaving \(\lambda\) fixed. Both strict inequalities are preserved. Finally, conditional on the constrained type, a run of \(K\) missed audits occurs with probability \((1-q)^K\), regardless of firm conduct. Weighting this event by its current probability \(\mu\) proves the lower bound.
\end{proof}

The bound \(K\) answers a conditional question: how many missed audits, starting from the specified history, are enough to guarantee the two actions regardless of the equilibrium played before entry? The lower probability bound shows that this finite history has positive probability. It does not show that the relationship will eventually reach the region with probability one. A finding can end the relationship, and a clean audit can raise the compliance belief enough to invalidate the starting restriction. Moreover, a particular equilibrium may prescribe the trap behavior before the sufficient bound is reached.

The simple bound does not cover arbitrarily patient players: \(\pi_\delta\) tends to zero and \(\beta_\delta\) eventually exceeds one. The next argument addresses this limitation by using more of the continuation structure. We retain the simple result because it can give a substantially shorter sufficient silence bound where its conditions hold.

\subsection{Entry for arbitrarily patient players}\label{sec:patient}

For patient players, the simple bounds allow a large difference between the inspector's continuation values after the two active signals. They do so even when the model restricts how different those continuations can be. We now use that restriction. If the capacity belief is sufficiently pessimistic, the strategic firm's conduct is determined over many future histories. The inspector then faces a known pattern of firm behavior for a long initial interval, and the effect of the remaining continuation can be bounded by discounting. This yields a sufficient result for every fixed \(\delta<1\), with a bound that need not remain useful as \(\delta\) approaches one.

For this result replace \eqref{eq:global-primitives} by
\begin{equation}\label{eq:patient-primitives}
q<p^*=\frac{g}{d(g+f)}<\ah,
\qquad c<d(r+\ell).
\end{equation}
Capacity separation continues to hold. The first restriction says that the constrained technology cannot provide one-period deterrence, whereas high functioning effort can. The second says that an additional audit has a positive one-period return when violation is certain. Neither restriction involves discounting.

\begin{lemma}\label{lem:firm-floor}
Under \eqref{eq:patient-primitives}, at every strategic-firm information set,
\begin{equation}\label{eq:firm-floor-value}
0\leq V\leq\overline V_\delta
=\frac{g-qd(g+f)}{1-\delta+\delta qd}.
\end{equation}
Violation is strictly optimal if the current audit probability satisfies
\begin{equation}\label{eq:patient-pi}
p<\widehat\pi_\delta
=\frac{g(1-\delta+\delta dq)}{d\{g+(1-\delta)f\}}.
\end{equation}
For every \(\delta\in(0,1)\), \(q<\widehat\pi_\delta<p^*\).
\end{lemma}

The proof is given in Appendix F.1. The positive audit floor is useful even though it is too low to provide one-period deterrence. A firm that violates repeatedly must continue to face a positive chance of detection and termination. Its future gains are therefore bounded more tightly than by allowing the payoff \(g\) in every remaining period. This sharper bound leaves a nonempty interval between \(q\) and \(\widehat\pi_\delta\) at every fixed discount factor below one.

The inspector comparison uses an auxiliary decision problem. A functioning inspector faces a committed firm with probability \(\lambda\), and otherwise a firm that violates in every period. In this problem violation is imposed, rather than asserted to be an equilibrium strategy everywhere. Let \(W^\circ(\lambda)\) be the inspector's optimal value. A missed audit leaves \(\lambda\) unchanged, and a clean audit updates it to \(\lambda/[1-d(1-\lambda)]\).

The values of always using low effort against the two firm types are
\[
w_C^L=-\frac{\al c}{1-\delta},\qquad
w_V^L=\frac{-\al c+\al d(r+\ell)-\ell}{1-\delta+\delta\al d}.
\]
Define
\begin{align}
T_\delta&=d(r+\ell-\delta w_V^L)
=\frac{d\{(1-\delta)r+\ell+\delta\al c\}}
{1-\delta+\delta\al d},\label{eq:aux-T}\\
a_\delta^\circ&=\frac{c}{T_\delta}
=\frac{c(1-\delta+\delta\al d)}
{d\{(1-\delta)r+\ell+\delta\al c\}}.\label{eq:aux-cutoff}
\end{align}

\begin{lemma}\label{lem:aux-control}
Under \eqref{eq:patient-primitives}, \(a_\delta^\circ\in(0,1)\). In the auxiliary problem, high effort is uniquely optimal if \(1-\lambda>a_\delta^\circ\), and low effort is uniquely optimal if \(1-\lambda<a_\delta^\circ\). Both are optimal at equality. If \(a=1-\lambda>a_\delta^\circ\), the gain per unit increase in the audit probability satisfies
\begin{equation}\label{eq:aux-margin}
\Psi^\circ(\lambda)\geq
\frac{1-\delta+\delta\al}{1-\delta+\delta\ah}
T_\delta(a-a_\delta^\circ)>0.
\end{equation}
\end{lemma}

The proof is given in Appendix F.1. In the auxiliary problem, a clean audit increases the chance of facing a committed firm, while no audit leaves that chance unchanged. When the probability of violation is low, low effort remains optimal after all subsequent active observations. When that probability exceeds the cutoff, the gain from high effort is positive by a known margin. The lower bound in \eqref{eq:aux-margin} lets us control how much the actual continuation problem can differ before high effort ceases to be optimal.

To apply the auxiliary comparison to the game, choose a compliance-belief cap \(\Lambda\in(0,1-a_\delta^\circ)\). This leaves a strictly positive margin for high effort in the auxiliary problem whenever the compliance belief is below the cap. The constants below keep track of that margin and of the possible error from ignoring distant continuation payoffs. In particular, \(\eta_\delta\) is a lower bound on the auxiliary gain per unit increase in the audit probability, \(C_\delta\) bounds the range of remaining payoffs, and \(O_F\) is the capacity-odds threshold above which the firm must violate. Set
\begin{align}
\eta_\delta&=
\frac{1-\delta+\delta\al}{1-\delta+\delta\ah}
T_\delta(1-\Lambda-a_\delta^\circ)>0,\label{eq:patient-margin}\\
C_\delta&=R+\frac{\ell+c}{1-\delta},\qquad
O_F=\frac{\ah-\widehat\pi_\delta}{\widehat\pi_\delta-q}.
\label{eq:patient-constants}
\end{align}
Choose an integer \(M\geq1\) satisfying
\begin{equation}\label{eq:patient-M}
2C_\delta\delta^{M+1}<\eta_\delta,
\end{equation}
and define
\begin{equation}\label{eq:patient-O}
O_I=O_F(\ah/q)^{M+1},\qquad
\Tset^P_\delta(\Lambda)=
\left\{(\lambda,\mu):\lambda<\Lambda,\ 
\frac{\mu}{1-\mu}>O_I\right\}.
\end{equation}
The role of \(M\) is to make the remaining discounted payoff difference smaller than the effort margin. The higher threshold \(O_I\) then ensures that the firm's capacity odds stay above \(O_F\) throughout the required finite set of active histories, even after clean audits. All constants are finite for each fixed \(\delta<1\). The region \(\Tset^P_\delta(\Lambda)\) supplies a second sufficient guarantee; its belief restrictions differ from those defining \(\Tset\).

\begin{theorem}\label{thm:patient}
Maintain Assumption~\ref{ass:separation} and \eqref{eq:patient-primitives}. For every \(\delta\in(0,1)\), every \(\Lambda\in(0,1-a_\delta^\circ)\), and \(M\) satisfying \eqref{eq:patient-M}, every history in \(\Tset^P_\delta(\Lambda)\) is a capacity-trap history in every sequential equilibrium.

Start at any active public history with \(\lambda<\Lambda\) and capacity odds \(O=\mu/(1-\mu)>0\). A run of
\begin{equation}\label{eq:patient-K}
K^P=\min\{k\in\mathbb Z_{\geq0}:OF^k>O_I\},
\qquad F=\frac{1-q}{1-\al},
\end{equation}
missed audits suffices for entry. The run has conditional probability at least \(\mu(1-q)^{K^P}>0\). Further missed audits preserve membership.
\end{theorem}

\begin{proof}
A capacity odds ratio larger than \(O_F\) makes even the maximum feasible audit probability smaller than \(\widehat\pi_\delta\). Lemma~\ref{lem:firm-floor} therefore forces every strategic firm to violate at every information set with that public history.

Under arbitrary equilibrium play, a missed audit multiplies capacity odds by at least \(F>1\); a clean audit multiplies them by \(q/x\geq q/\ah\). Consequently, from a history with odds \(O>O_I\), every active public successor at depth at most \(M+1\) has odds larger than \(O_F\). Strategic firms violate throughout this finite collection of histories, including at private information sets reached after deviations.

Fix the functioning inspector's information set at the starting history. At either active one-period successor, its first \(M\) continuation periods coincide with the auxiliary problem: committed firms comply, strategic firms violate, and the signal and payoff kernels agree. This agreement holds under any inspector continuation plan. Every such plan has a remaining payoff between \(-(\ell+c)/(1-\delta)\) and \(R\). Coupling the same plan for the first \(M\) periods and then allowing arbitrary tails thus bounds the difference between the optimal continuation values by \(C_\delta\delta^M\). The private-history result of Lemma~\ref{lem:beliefs} supplies the same initial opponent distribution at each inspector information set. The Online Appendix details this finite-horizon comparison.

Using \eqref{eq:inspector-diff}, the difference between the actual and auxiliary effort gains is at most
\[
\delta\{(1-da)+1\}C_\delta\delta^M
\leq2C_\delta\delta^{M+1},\qquad a=1-\lambda.
\]
The auxiliary gain is at least \(\eta_\delta\) whenever \(\lambda\leq\Lambda\), by Lemma~\ref{lem:aux-control}. Condition~\eqref{eq:patient-M} therefore makes high effort strictly optimal in the game. Firm violation was already strict, and the resulting audit probability is below \(\widehat\pi_\delta<p^*\).

Finally, the likelihood-ratio argument in Theorem~\ref{thm:global} gives \(O_{K^P}\geq OF^{K^P}>O_I\), with unchanged \(\lambda\). Another missed audit increases the odds and preserves the two conditions. Conditioning on constrained capacity proves the stated probability bound.
\end{proof}

The compliance-prior requirement can also be made independent of discounting. The expression \(a_\delta^\circ\) is a ratio of positive affine functions of \(\delta\), so its maximum over \([0,1]\) occurs at an endpoint. Hence a fixed initial belief satisfying
\begin{equation}\label{eq:patient-fixed-prior}
\lambda_0<1-\max\left\{
\frac{c}{d(r+\ell)},\frac{\al c}{\ell+\al c}\right\}
\end{equation}
admits a single cap \(\Lambda>\lambda_0\) for all \(\delta<1\). The silence bound still depends on \(\delta\). This is a family of discounted-game results, not a theorem at \(\delta=1\) or a uniform bound in that limit.

\begin{example}\label{ex:patient}
Use the payoff and technology parameters
\[
(g,f,c,r,\ell,d,q,\al,\ah)=(1,4,0.3,1,1,0.9,0.02,0.2,0.8).
\]
Start at \((\lambda_0,\mu_0)=(0.05,0.5)\) and choose \(\Lambda=0.1\). The smallest \(M\geq1\) satisfying \eqref{eq:patient-M} gives:
\begin{center}
\begin{tabular}{@{}rrrrr@{}}
\toprule
\(\delta\)&\(\widehat\pi_\delta\)&\(a_\delta^\circ\)&\(M\)&\(K^P\)\\
\midrule
0.50&0.18852&0.12854&2&61\\
0.90&0.09222&0.07568&30&575\\
0.95&0.06213&0.06655&75&1,396\\
0.99&0.02972&0.05866&542&9,892\\
\bottomrule
\end{tabular}
\end{center}
These are sufficient bounds, not computed equilibrium entry times. For \(\delta=0.9\), the probability lower bound is approximately \(4.51\times10^{-6}\); for \(\delta=0.99\), it is approximately \(8.08\times10^{-88}\). A small lower bound does not show that actual entry is rare. It shows that this proof cannot establish a useful frequency guarantee. Where the simpler theorem applies, its bound can be much shorter.
\end{example}

The extension preserves the economic restriction on capacity. It improves the incentive argument, rather than relaxing the assumption that silence uniformly favors the constrained type. More generally, along a particular silence path the exact identity is
\[
\log O_n=\log O_0+
\sum_{t=0}^{n-1}\log\frac{1-q}{1-x_t}.
\]
A diverging positive sum would also give entry into the patient-player region, but without separation that property requires an additional equilibrium argument. At \(q=\al\), low effort makes its summands zero; at \(q>\al\), they are negative under low effort. These observations identify where the proof fails. They are not counterexamples establishing what equilibrium must do when capacity ranges overlap.

\subsection{Clean audits, departure, and re-entry}\label{sec:clean-exit}

A missed audit leaves the firm's reputation unchanged. A clean audit provides evidence about both parties, and that difference determines whether the preceding result can continue to apply. Starting inside \(\Tset\), write \(p=p^H(\mu)\). Since all strategic firms violate, the successor beliefs are
\begin{equation}\label{eq:clean-in-trap}
\lambda_A=\frac{\lambda}{1-d(1-\lambda)},\qquad
\mu_A=\frac{\mu q}{\mu q+(1-\mu)\ah}.
\end{equation}
The first posterior rises and the second falls. A clean audit suggests that the firm may be committed to compliance and that the inspector may possess functioning capacity. Both changes weaken the sufficient conditions for the configuration of universal strategic violation and high functioning effort.

\begin{proposition}\label{prop:clean-exit}
Under the hypotheses of Theorem~\ref{thm:global}, start at a history in \(\Tset\). After a clean audit the successor remains in \(\Tset\) if and only if
\begin{align}
\lambda&<\lambda^A_\delta
\equiv\frac{(1-\beta_\delta)(1-d)}{1-d+d\beta_\delta},\label{eq:lambda-clean-cutoff}\\
\mu&>\mu^A_\delta
\equiv\frac{\mu^\dagger\ah}
{q(1-\mu^\dagger)+\mu^\dagger\ah}.\label{eq:mu-clean-cutoff}
\end{align}
The new cutoffs satisfy \(0<\lambda^A_\delta<1-\beta_\delta\) and \(\mu^\dagger<\mu^A_\delta<1\). If either strict inequality fails, the successor is outside the sufficient region, including when equality holds.
\end{proposition}

\begin{proof}
Apply Lemma~\ref{lem:beliefs} with \(a=1-\lambda\) and \(x=\ah\), obtaining \eqref{eq:clean-in-trap}. All denominators are positive. Rearranging \(\lambda_A<1-\beta_\delta\) gives
\[
\lambda\{1-d+d\beta_\delta\}<(1-\beta_\delta)(1-d).
\]
Rearranging \(\mu_A>\mu^\dagger\) gives
\[
\mu\{q(1-\mu^\dagger)+\mu^\dagger\ah\}>\mu^\dagger\ah.
\]
These are exactly the two displayed conditions. The cutoff orderings follow from \(d,\beta_\delta,\mu^\dagger\in(0,1)\) and \(q<\ah\).
\end{proof}

The proposition identifies departure from a set of sufficient conditions. It does not imply that compliance or low effort is optimal immediately after departure. Those choices depend on continuation values that the bounds do not determine. The distinction is important: the posterior calculation is exact, while the region itself is a sufficient characterization.

\begin{corollary}\label{cor:clean-reentry}
If a clean audit from \(\Tset\) yields \(\lambda_A<1-\beta_\delta\) but \(\mu_A\leq\mu^\dagger\), a sufficiently long subsequent uninterrupted run of missed audits returns beliefs to \(\Tset\) in every sequential equilibrium. The return run has positive conditional probability, with its sufficient length obtained from \eqref{eq:K} using \(\mu_A\).

If instead \(\lambda_A\geq1-\beta_\delta\), no later active public history can belong to this same sufficient region.
\end{corollary}

\begin{proof}
The first case satisfies the compliance-belief hypothesis of Theorem~\ref{thm:global} at the clean-audit successor. In the second case, Lemma~\ref{lem:beliefs} implies that every subsequent missed audit leaves \(\lambda\) unchanged and every clean audit weakly increases it. Thus the strict compliance-belief condition can never be restored.
\end{proof}

The two cases describe different kinds of learning. A clean audit can restore confidence that the inspector has functioning capacity, and that confidence can later be eroded by another run of silence. Evidence of committed compliance has a different effect: its posterior never falls along an active history. Once it exceeds the cutoff of the chosen region, that particular sufficient guarantee is lost permanently. This still leaves open whether the equilibrium may prescribe capacity-trap behavior at other histories outside the region.

\begin{proposition}\label{prop:finite-clean-exit}
The sufficient region \(\Tset\) is not absorbing. From any history in it, there is a finite sequence of clean audits, with positive conditional probability, whose last observation first takes beliefs outside \(\Tset\).
\end{proposition}

\begin{proof}
Write the starting compliance and capacity odds as \(L=\lambda/(1-\lambda)\) and \(O=\mu/(1-\mu)\). As long as each preceding history remains in \(\Tset\), a clean audit multiplies these odds by \((1-d)^{-1}\) and \(q/\ah\), respectively. Hence after \(n\) consecutive clean audits, up to and including the first departure,
\[
L_n=L(1-d)^{-n},\qquad O_n=O(q/\ah)^n.
\]
The first departure occurs at
\[
m=\min\left\{n\geq1:
L(1-d)^{-n}\geq\frac{1-\beta_\delta}{\beta_\delta}
\ \text{or}\ O(q/\ah)^n\leq O^\dagger\right\}.
\]
This integer is finite, since \(L>0\), \(d>0\), and \(q/\ah<1\). Before that observation both actions are fixed by dominance. The probability of the specified \(m\) clean audits is therefore
\[
\{\lambda+(1-\lambda)(1-d)^m\}
\{\mu q^m+(1-\mu)\ah^m\}>0.
\]
The product follows by conditioning on the initially independent types at the starting public history. A clean audit is nonterminal, so the last observation is an active departure rather than termination.
\end{proof}

One clean audit can leave a sufficiently deep history inside the region. A finite run of clean audits nevertheless produces departure. Conversely, any finite run consisting only of missed audits preserves membership once it is reached. These observations explain the sense in which the configuration is a trap sustained by silence without being an absorbing state of the game.

\paragraph{The patient-player region.}
For a fixed cap \(\Lambda\) and window \(M\), the region in Theorem~\ref{thm:patient} is also a rectangle. Set \(\mu_I=O_I/(1+O_I)\). All the preceding clean-audit calculations apply to it after replacing \(1-\beta_\delta\) by \(\Lambda\) and \(\mu^\dagger\) by \(\mu_I\). In particular, a clean audit preserves membership exactly when
\[
\lambda<\frac{\Lambda(1-d)}{1-d\Lambda},\qquad
\mu>\frac{\mu_I\ah}{q(1-\mu_I)+\mu_I\ah}.
\]
Leaving through the capacity belief permits re-entry after silence. Crossing the chosen compliance cap prevents return to that particular rectangle. It need not invalidate every patient-player guarantee, since a larger cap can still be admissible if \(\lambda_A<1-a_\delta^\circ\). The Online Appendix records the corresponding finite-clean-audit departure calculation.

\subsection{Enforcement and harm after further silence}

Inside either sufficient region, all strategic firms violate and functioning inspectors use high effort. Write \(a=1-\lambda\) and \(p(\mu)=\mu q+(1-\mu)\ah\). Another no-audit observation leaves \(\lambda\) unchanged, raises \(\mu\), and preserves the region. It therefore lowers \(p\), while neither type's implementation probability changes.

\begin{corollary}\label{cor:harm}
At such a history, the detection and termination hazard is \(\chi=dap\), and the probability of an undetected violation is \(u=a(1-dp)\). After another no-audit observation, \(\chi\) strictly falls and \(u\) strictly rises.
\end{corollary}

\begin{proof}
Both expressions follow by conditioning on an audit and a violation. The no-audit update lowers \(p\) and leaves \(a>0\) fixed.
\end{proof}

For a one-period social-loss comparison, let \(\mathcal H>0\) denote undetected harm, \(C_s\geq0\) the real cost of an audit, and \(K_0\geq0\) the real cost of termination. Fines and compensation are transfers. Expected current loss is
\begin{equation}\label{eq:social-loss}
\mathcal L=\mathcal H a(1-dp)+C_s p+K_0dap.
\end{equation}
Another missed audit raises this expected loss if \(da(\mathcal H-K_0)>C_s\). This is a comparison of next-period expected losses along a specified public-history path. The following section separately compares losses across technologies over a common horizon.

\section{Comparative statics of enforcement capacity}\label{sec:design}

Consider two enforcement systems with the same initial beliefs and the same audit probability if the functioning inspector uses high effort. They can nevertheless differ in the dispersion of their capacities. One may combine a very low constrained rate with a high functioning rate; the other may provide a better floor but a lower peak. We ask how this difference affects learning from silence and, where behavior can be determined, expected social loss. The comparison holds the initial mean fixed. Whether it also holds the resources needed to build the systems fixed depends on the cost specification.

\subsection{Capacity dispersion and learning}

Let \(\Theta\in(0,1)\) be a persistent audit propensity under a specified, unchanged effort rule, with current distribution \(G\), mean \(p\), and variance \(\nu\). For example, under high functioning effort in the binary model, \(\Theta=q\) with probability \(\mu\) and \(\Theta=\ah\) otherwise. The following identities concern the posterior of that propensity. They become predictions about the next actual audit probability only if the effort rule continues to apply.

\begin{lemma}\label{lem:design}
For any such distribution,
\begin{align}
\E[\Theta\mid\No]&=p-\frac{\nu}{1-p},\label{eq:pN-general}\\
\E[\Theta\mid\text{audit}]&=p+\frac{\nu}{p}.\label{eq:pA-general}
\end{align}
The audit identity also applies after a clean audit if the additional no-finding likelihood is the same at every capacity state. This condition follows from the model's product-form filter when firm behavior is held fixed.
\end{lemma}

\begin{proof}
Bayes' rule weights the capacity distribution by \(1-\theta\) after no audit and by \(\theta\) after an audit. The two posterior means are therefore \(\E[\Theta(1-\Theta)]/(1-p)\) and \(\E[\Theta^2]/p\). Substitution of \(\E[\Theta^2]=p^2+\nu\) gives the identities. A common no-finding factor cancels in the clean-audit update.
\end{proof}

Greater dispersion makes audit incidence more informative about capacity. A missed audit then lowers the posterior mean by more, and an audit raises it by more. Reducing dispersion moderates both responses. The mean before observing either outcome remains \(p\), so this calculation alone gives no increase in expected enforcement before the signal is realized. To interpret a posterior mean as next-period auditing, we must also establish that the effort rule used to define the propensity continues to apply. A clean audit may prevent this by taking beliefs outside the sufficient region.

In the binary model, fix the initial capacity belief \(\mu\) and maximum-effort mean \(p\). A reduction in the spread \(s=\ah-q\) can be written as
\begin{equation}\label{eq:spread-parameterization}
q(s)=p-(1-\mu)s,\qquad \ah(s)=p+\mu s.
\end{equation}
We compare feasible spreads with \(0<q(s)<\al<\ah(s)<1\), keeping \(\al\), payoffs, discounting, and the initial compliance belief fixed. A smaller spread raises the floor and lowers the peak. Equivalently,
\begin{equation}\label{eq:compression}
q'=q+\varepsilon,\qquad \ah'=\ah-\frac{\mu}{1-\mu}\varepsilon,
\end{equation}
where \(\varepsilon>0\) preserves the ordering. The variance falls from \(\mu(1-\mu)(\ah-q)^2\) to \(\mu(1-\mu)(\ah'-q')^2\).

To interpret the resource comparison, suppose creating audit propensity \(z\) has cost \(k(z)\). With a common linear cost, the two systems have the same expected capacity cost because their means agree. A common convex cost weakly favors compression. With different cost functions for constrained and functioning capacity, however, the local change in expected construction cost is
\[
\mu\{k_C'(q)-k_F'(\ah)\}\varepsilon+o(\varepsilon),
\]
whose sign need not be determined. A possible institutional interpretation is to provide implementation resources in the constrained state---for example, access to shared administrative support---while reducing resources available at peak functioning capacity. In the model, that proposal is represented by raising \(q\) and lowering \(\ah\). Whether a particular reform can implement this pair of changes at the stated costs is a separate institutional question. These are explicit alternative resource assumptions. The model does not supply an optimal allocation without choosing among them and accounting for subsequent equilibrium behavior.

\subsection{Silence paths and sufficient entry bounds}

Suppose high functioning effort is used during the first \(k\) periods, all of which produce no audit. The posterior capacity belief and maximum feasible next audit rate are
\begin{align}
\mu_k(s)&=\frac{\mu[1-q(s)]^k}{\mu[1-q(s)]^k+(1-\mu)[1-\ah(s)]^k},\label{eq:muk}\\
p_k^H(s)&=\mu_k(s)q(s)+[1-\mu_k(s)]\ah(s).\label{eq:pk}
\end{align}
The latter is the actual next audit rate if high effort continues. Inside a sufficient dominance region, continuation after further silence justifies that behavior in every equilibrium. Before entry, it is a specified effort path, rather than a characterization of equilibrium.

\begin{proposition}\label{prop:silence-path}
For feasible spreads \(s'<s\),
\[
p_k^H(s')>p_k^H(s)\qquad\text{for every integer }k\geq1.
\]
Consequently, for any \emph{common} audit threshold \(\tau\), the first index at which \(p_k^H<\tau\) is weakly later under \(s'\), with an empty crossing set assigned infinity.
\end{proposition}

\begin{proof}
For \(k\geq1\), \(\mu_k(s)>\mu\). Moreover,
\[
\frac{d}{ds}\log\frac{1-q(s)}{1-\ah(s)}
=\frac{1-\mu}{1-p+(1-\mu)s}+\frac{\mu}{1-p-\mu s}>0,
\]
so \(\mu_k'(s)>0\). Since \(p_k^H(s)=p+s[\mu-\mu_k(s)]\), its derivative is strictly negative. The common-threshold statement follows pointwise.
\end{proof}

This proposition applies, for example, to the one-period threshold \(p^*\) and the original sufficient threshold \(\pi_\delta\), which do not change with \(q\). It does not directly order crossings of the patient-player threshold \(\widehat\pi_\delta\), which does depend on \(q\).

The original entry theorem also permits an ordering that leaves pre-entry effort unrestricted. Suppose \(p>\pi_\delta\), \(q(s)<\pi_\delta<\ah(s)\), and \(\lambda<1-\beta_\delta\). Define
\begin{align}
\mu^\dagger(s)&=\mu+\frac{p-\pi_\delta}{s},\label{eq:mudagger-s}\\
F(s)&=\frac{1-q(s)}{1-\al},\label{eq:F-s}\\
K^U(s)&=\min\left\{k\geq0:
\frac{\mu}{1-\mu}F(s)^k>
\frac{\mu^\dagger(s)}{1-\mu^\dagger(s)}\right\}.\label{eq:KU}
\end{align}

\begin{proposition}\label{prop:floor-bound}
Under these conditions, \(K^U(s)\) missed audits suffice for entry in every sequential equilibrium, and \(s'<s\) implies \(K^U(s')\geq K^U(s)\).
\end{proposition}

\begin{proof}
Theorem~\ref{thm:global} supplies the bound. Compression raises \(\mu^\dagger(s)\) and lowers \(F(s)\), so every integer satisfying the entry inequality under the compressed technology also satisfies it under the original technology.
\end{proof}

The ordering concerns the bound delivered by the simple continuation comparisons. Compression both raises the posterior target in that argument and slows the minimum learning rate toward it. Those two changes explain the longer bound. Actual equilibrium entry also depends on behavior at beliefs where the sufficient comparisons are inconclusive. The patient-player construction presents a separate issue: changing capacity alters the firm's sharper threshold and the inspector's continuation comparison, so the same ordering need not apply to that bound either.

The numerical comparisons in Appendix G distinguish these two bounds. In particular, they show that compression can shorten the patient-player bound even though it lengthens the original bound. This limits the interpretation of the comparative static: a change in a sufficient bound is not an ordering of equilibrium entry times.

\subsection{Welfare over a common horizon}\label{sec:fixed-welfare}

We next compare losses over the same dates under both technologies. For this purpose, knowing what happens after further silence is not enough. A relationship can also remain active after a clean audit, and that observation may change the players' behavior. The comparison must therefore account for every active history over the horizon being used. The next lemma gives a sufficient condition: start far enough inside a dominance region that even a sequence of clean audits cannot take beliefs outside it before the last decision included in the calculation.

Write either sufficient region from Section~\ref{sec:global} as
\[
\mathcal R=\{(\lambda,\mu):\lambda<\Lambda,\quad
O\equiv\mu/(1-\mu)>\overline O\}.
\]
For the original theorem, \(\Lambda=1-\beta_\delta\) and \(\overline O=O^\dagger\). For the patient-player theorem, use its chosen cap and \(\overline O=O_I\).

\begin{lemma}\label{lem:finite-window}
If, for an integer \(n\geq1\),
\begin{equation}\label{eq:window-guard}
\frac{\lambda}{1-\lambda}(1-d)^{-(n-1)}
<\frac{\Lambda}{1-\Lambda},\qquad
O(q/\ah)^{n-1}>\overline O,
\end{equation}
then every active history before decisions in periods \(0,\ldots,n-1\) remains in \(\mathcal R\). Strategic violation and high functioning effort are therefore strict in every equilibrium throughout those periods.
\end{lemma}

\begin{proof}
At a history in the region, a clean audit multiplies compliance odds by \((1-d)^{-1}\) and capacity odds by \(q/\ah\). No audit leaves compliance odds unchanged and raises capacity odds. At every depth at most \(n-1\), the all-clean history thus gives the largest compliance odds and smallest capacity odds. The strict inequalities allow induction over the whole active history tree.
\end{proof}

The exponent is \(n-1\) because the starting history already supplies the first decision in the horizon. The condition must keep beliefs in the region before each of the remaining \(n-1\) decisions. Beliefs after the last period may leave the region, since the comparison does not use the actions that would then follow.

Consider technologies satisfying this condition at the same initial beliefs and the same horizon \(n\). Let \(\mathcal H>0\) be real harm from an undetected violation, \(C_s\geq0\) an audit's resource cost, and \(K_0\geq0\) a real cost of detection-induced termination. Fines and compensation are transfers. These social costs are additional primitives; their values cannot be inferred from the players' private payoffs. Put
\begin{equation}\label{eq:welfare-B}
B=C_s+d(K_0-\mathcal H).
\end{equation}
Conditional on a strategic firm and propensity \(\theta\), current expected loss is \(\mathcal H+B\theta\), and survival has probability \(1-d\theta\). A committed firm instead incurs expected audit cost \(C_s\theta\) each period and always survives. Thus expected discounted loss through the common horizon is the average over capacity of
\begin{equation}\label{eq:fixed-welfare}
J_n^{\mathrm{pop}}(\theta)
=\lambda C_s\theta\sum_{t=0}^{n-1}\delta^t
+(1-\lambda)(\mathcal H+B\theta)
\sum_{t=0}^{n-1}[\delta(1-d\theta)]^t.
\end{equation}
The first term in \eqref{eq:fixed-welfare} is the contribution of committed firms. They always survive, so their audit costs are counted at every date. The second term is the contribution of strategic firms. Its geometric factor includes both discounting and the probability that detection has not yet ended the relationship. The initial type weights therefore remain \(\lambda\) and \(1-\lambda\): selection is already included in the survival factor. Both clean audits and missed audits are included, and losses are counted until detection or the common date \(n\), whichever comes first.

\begin{proposition}\label{prop:welfare}
Suppose \(n\geq2\), both technologies satisfy Lemma~\ref{lem:finite-window}, and
\begin{equation}\label{eq:welfare-cond}
d(\mathcal H-K_0)>C_s.
\end{equation}
A strict fixed-mean capacity compression lowers expected discounted social loss over the common horizon in every sequential equilibrium of either technology. More generally, the loss kernel \(J_n^{\mathrm{pop}}\) is strictly convex in \(\theta\), so the same mathematical comparison holds for any nontrivial mean-preserving contraction whenever the stipulated behavior is justified.
\end{proposition}

\begin{proof}
Condition~\eqref{eq:welfare-cond} says \(B<0\). The committed-firm term is linear in \(\theta\). For the strategic term at date \(t\geq1\), differentiation gives
\[
\frac{d^2}{d\theta^2}\left[\delta^t(1-d\theta)^t(\mathcal H+B\theta)\right]
=\delta^t td(1-d\theta)^{t-2}
\left[(t-1)d(\mathcal H+B\theta)-2B(1-d\theta)\right]>0.
\]
Here \(\mathcal H+B\theta=\mathcal H(1-d\theta)+C_s\theta+K_0d\theta>0\). The date-zero term is linear, and the date-one term is already strictly convex. Since \(\lambda<1\), averaging the kernel under a strict mean-preserving contraction lowers loss. Lemma~\ref{lem:finite-window} makes the behavior used in the calculation valid in every equilibrium.
\end{proof}

The economic reason for the convexity is that low capacity affects both the current loss and the duration over which losses can occur. Under \eqref{eq:welfare-cond}, a higher audit probability reduces the current expected loss of a strategic relationship. It also raises the chance of detecting the violation and ending that relationship before another loss occurs. Persistent low-capacity relationships consequently contribute disproportionately to expected loss over the horizon. A compression reduces this dispersion while leaving the initial mean unchanged.\footnote{The general statement about mean-preserving contractions concerns the loss function under the stipulated behavior. The equilibrium model developed here has two inspector types. Applying the same calculation to a richer distribution of capacities would also require a justification of behavior under that specification.}

For \(n=2\), the improvement has a particularly simple form. If the initial capacity variance falls from \(\nu\) to \(\nu'\), then
\begin{equation}\label{eq:two-period-welfare-gain}
\E_G[J_2^{\mathrm{pop}}]-\E_{G'}[J_2^{\mathrm{pop}}]
=(1-\lambda)\delta d(-B)(\nu-\nu')>0.
\end{equation}
The sign condition matters: when \(B>0\), the two-period comparison reverses. The result also leaves losses after the common horizon unranked. It is not a lifetime welfare theorem, nor a claim about unmodeled construction costs.

\begin{example}\label{ex:welfare}
Let \((g,f,c,r,\ell,d,\delta)=(1,4,0.1,1,1,0.9,0.5)\), \(\al=0.3\), and \((\lambda,\mu)=(0.01,0.98)\). Compare
\[
(q,\ah)=(0.1,0.7),\qquad
(q',\ah')=(0.10612245,0.4),
\]
where \(q'=0.1+(0.02/0.98)0.3\) exactly. Both initial means are \(0.112\). The original region has compliance cutoff \(1-\beta_\delta=0.4222222\); its capacity cutoffs are \(0.8580247\) and \(0.7309671\). After the worst-case one-period clean audit, the compliance belief is \(0.0917431\), and capacity beliefs are \(0.875\) and \(0.9285714\). Both technologies therefore satisfy \eqref{eq:window-guard} for \(n=2\).

For \((\mathcal H,C_s,K_0)=(10,0.2,0)\), \(B=-8.8\). Expected first-period loss is \(8.92448\) under both systems. Expected loss over two periods is \(12.96460\) under the original technology and \(12.94357\) after compression. The improvement, \(0.0210261\), agrees with \eqref{eq:two-period-welfare-gain}. These calculations use the same discount factor as the equilibrium conditions.
\end{example}

The comparison through the first audit is a valid, differently stopped calculation when both systems start in a sufficient region. Its derivation and an illustration using these same admissible parameters are retained in the Online Appendix. The common-horizon result above does not rely on that stopping convention.

\section{Selection among surviving relationships}\label{sec:survivors}

Suppose an observer knows that a relationship is still active but does not know its complete audit history. Survival can have several explanations. The firm may comply, in which case detection never ends the relationship. Alternatively, a strategic firm may violate and remain active because the inspector rarely implements an audit or because a violation goes undetected. These possibilities create selection across relationships. We first examine one period and then study duration over a horizon on which the preceding incentive conditions determine behavior.

Let \(H=1\) denote a committed firm and \(P=1\) a constrained inspector. At a fixed active public history the types are independent, with probabilities \(\lambda\) and \(\mu\). Let the strategic firm's current violation probability be \(\sigma\), the functioning audit rate be \(x\), and
\[
a=(1-\lambda)\sigma,\qquad p=\mu q+(1-\mu)x,
\qquad D=1-dap.
\]

\begin{proposition}\label{prop:sorting}
If an observer learns only that the relationship survives the current period, then
\begin{align}
\lambda_S&=\frac{\lambda}{D},\label{eq:lambdaS}\\
\mu_S&=\frac{\mu(1-daq)}{D},\label{eq:muS}\\
\Cov(H,P\mid S)&=\frac{\lambda\mu da(q-p)}{D^2}.\label{eq:covS}
\end{align}
When \(a>0\) and \(q<p\), both marginal probabilities rise and the covariance is strictly negative. With \(a=0\), neither marginal changes and the covariance is zero.
\end{proposition}

\begin{proof}
A committed firm always survives. A constrained inspector survives with probability \(1-daq\), averaging over firm types. The joint event of committed compliance and constrained capacity always survives and has probability \(\lambda\mu\). Dividing these probabilities by \(D\) gives the marginals and joint posterior; subtracting their product gives the covariance.
\end{proof}

Inside either sufficient region, \(\sigma=1\) and \(x=\ah\), so the proposition applies to the behavior already established. Its two marginal changes have straightforward explanations. Committed firms survive with certainty, and constrained inspectors are less likely to detect a violation. Both types therefore become more prevalent in the surviving sample. Their negative association concerns how they are paired: among surviving strategic firms, constrained inspectors are relatively common, whereas a committed firm survives regardless of its inspector's type. These facts alone do not determine the strategic/constrained pair's share of all survivors.

To examine duration, suppose Lemma~\ref{lem:finite-window} holds for \(n\) periods under the technology being studied. For \(0\leq t\leq n\), write
\begin{equation}\label{eq:survival-Z}
A_q=1-dq,\qquad A_{\ah}=1-dh,\qquad
Z_t=\lambda+(1-\lambda)\{\mu A_q^t+(1-\mu)A_{\ah}^t\}.
\end{equation}
The expression \(Z_t\) is the probability that the relationship survives for \(t\) periods. Committed firms contribute the constant term \(\lambda\), since they never produce a finding. A strategic firm survives each period with probability \(A_q\) when paired with constrained capacity and with probability \(A_{\ah}\) when paired with functioning capacity using high effort. The powers in \eqref{eq:survival-Z} average over all possible nonterminal audit records. The finite-window condition matters because it justifies the same actions on every one of those records.

\begin{proposition}\label{prop:duration}
Under the finite-window conditions, among surviving strategic firms the constrained-inspector share is
\begin{equation}\label{eq:strategic-survivor-share}
m_t=\frac{\mu A_q^t}{\mu A_q^t+(1-\mu)A_{\ah}^t}.
\end{equation}
It strictly increases with \(t\) on the stated horizon. The share of the strategic/constrained pair among \emph{all} survivors is instead
\begin{equation}\label{eq:harmful-survivor-share}
b_t=\frac{(1-\lambda)\mu A_q^t}{Z_t}.
\end{equation}
For \(t<n\), it rises from \(t\) to \(t+1\) if and only if
\begin{equation}\label{eq:harmful-rise}
(1-\lambda)(1-\mu)(\ah-q)A_{\ah}^t>\lambda q.
\end{equation}
Equality gives no change. Once this inequality fails, it cannot become true later within the same justified horizon. The survivor-sample odds ratio is
\begin{equation}\label{eq:OR}
\operatorname{OR}_t=(A_{\ah}/A_q)^t<1\qquad(t\geq1).
\end{equation}
\end{proposition}

\begin{proof}
The four unnormalized type-pair weights after survival are
\[
\lambda\mu,\quad\lambda(1-\mu),\quad
(1-\lambda)\mu A_q^t,\quad(1-\lambda)(1-\mu)A_{\ah}^t.
\]
Normalizing them gives the two shares and the odds ratio. Since \(A_q>A_{\ah}\), the constrained share among strategic survivors rises. For the population share, comparison of \(b_{t+1}\) and \(b_t\) reduces to \(A_qZ_t>Z_{t+1}\). Expanding and canceling the constrained-strategic term gives \eqref{eq:harmful-rise}. Its left side decreases with \(t\).
\end{proof}

The distinction between \(m_t\) and \(b_t\) comes from their denominators. The first compares only strategic firms: violations are detected more frequently under functioning capacity, so constrained inspectors account for an increasing share of the strategic firms that remain. The second also includes committed firms, all of which survive. A harmful pair can therefore gain relative to other strategic relationships while losing relative to the whole surviving population. In \eqref{eq:harmful-rise}, the left side measures the selection against functioning-strategic relationships, while the right side reflects the presence of committed firms and the positive detection risk faced by a constrained-strategic pair.

\begin{example}\label{ex:survivors}
Under the original technology in Example~\ref{ex:welfare}, the harmful-pair shares at durations zero, one, and two are \(0.97020\), \(0.98075\), and \(0.98443\). The corresponding constrained shares among strategic survivors are \(0.98000\), \(0.99177\), and \(0.99664\). These two-period calculations satisfy the finite-window conditions.

Alternatively, keep that technology and all payoffs but change the initial compliance belief to \(\lambda=0.2\). The starting history is still inside the original sufficient region, so the one-period calculation remains an every-equilibrium implication. The harmful-pair share then falls from \(0.78400\) to \(0.77602\), even though the constrained share among strategic survivors rises from \(0.98\) to \(0.99177\). Thus the stronger population claim can fail at a trap history itself.
\end{example}

The full no-audit record gives a different comparison. Once either sufficient region is reached, any number of further missed audits preserves the actions. Conditional on precisely that record,
\begin{equation}\label{eq:noaudit-harmful-share}
\Prob(H=0,P=1\mid\No^t)
=(1-\lambda)
\frac{\mu(1-q)^t}{\mu(1-q)^t+(1-\mu)(1-\ah)^t},
\end{equation}
which strictly increases with \(t\), while the two types remain independent. In this calculation, the observation is precisely a sequence of missed audits. That sequence reveals capacity but gives no direct information about firm conduct, leaving the compliance belief unchanged. Survival alone contains less information and also permits clean audits, so its implications for the composition of the sample differ.

The finite-duration formulas cannot in general be extrapolated indefinitely in equilibrium: clean audits may eventually leave the sufficient region. As a diagnostic, if one imposes violation and high effort forever, then \(m_t\to1\) but \(b_t\to0\) whenever \(\lambda>0\) and \(q>0\). The latter limit follows because committed firms always survive while both strategic pairings eventually exit. These are limits of a fixed-behavior benchmark, not long-run equilibrium predictions.

For an empirical application of the pooled-survival result, the analyst must observe duration while lacking or deliberately discarding some of the audit record. Administrative data that record closures or findings but omit audits without findings are one possible information structure. A study with the complete audit record should instead condition on that record, where the product-form posterior continues to apply. Likewise, applying the formulas to a sample with heterogeneous starting beliefs or other exit risks requires modeling those additional sources of selection.

\section{Implications and scope}\label{sec:implications}

The distinction between effort and implementation is useful when interpreting a low audit rate. Within either sufficient region, the functioning inspector is already choosing high effort. Further silence lowers the audit probability expected from public information because the constrained type becomes more likely. An empirical interpretation would therefore require evidence on the opportunities to implement audits as well as on their observed frequency. The separation between detection and enforcement capacity emphasized by \citet{Okunogbe2021} is relevant to this measurement problem. The present model additionally requires the specified audit record and a finding that ends, or resets, the relationship.

The capacity comparison asks what changes when a system provides a better floor at the expense of its peak. Holding the initial maximum-effort mean fixed isolates the effect of persistent capacity dispersion. It does not establish that the change is costless: a common linear construction cost makes the comparison budget neutral in expectation, a common convex cost favors compression, and different state-specific costs can give another ranking. The common-horizon welfare result concerns losses generated by auditing, undetected violations, and termination. A complete policy evaluation would also have to specify construction costs and equilibrium behavior beyond that horizon.

Duration introduces an additional distinction. A declining detection hazard can reflect learning along a known audit history, but it can also reflect selection across relationships with different persistent hazards. The survivor calculations make the observer's information explicit. They show why negative association between compliance and constrained capacity does not by itself imply an increasing harmful share in the whole population. Applying these calculations to heterogeneous cohorts or to data with other reasons for exit would require those additional sources of selection to be modeled.

The main theorem uses capacity separation to make the direction of learning independent of equilibrium effort. When the capacity ranges overlap, a missed audit need not favor constrained capacity; the learning argument then requires additional restrictions on play. The patient-player result retains this assumption. Its improvement is to the incentive comparison, which works at every fixed discount factor below one. The examples also show its quantitative limitation: the sufficient silence length can grow rapidly with patience, and the theorem does not establish that entry is likely.

\section{Conclusion}\label{sec:conclusion}

This paper studies a repeated enforcement relationship in which an inspector's effort and its ability to implement an audit are distinct. A firm can interpret repeated absence of audits as evidence of constrained capacity. Under the stated payoff and belief conditions, sufficiently long silence then makes violation strictly optimal even though a functioning inspector exerts maximum effort. The result applies to every sequential equilibrium at the specified histories, for each fixed discount factor below one. Clean audits convey different information and can take beliefs outside the region on which the argument determines behavior.

The capacity and welfare results explain one consequence of persistent differences in implementation. Two systems with the same initial maximum-effort mean can respond differently to silence because their capacities are distributed differently. A higher floor sustains a higher audit ceiling along the specified silence paths. When both systems remain in a sufficient region over a common finite horizon, a reduction in dispersion also lowers expected loss under the stated social-cost condition. The finite-horizon restriction allows that conclusion without assuming that all future clean audits leave behavior unchanged.

Survival changes the information available to an observer as well as the composition of a population. Among strategic firms that remain active, constrained inspectors become more prevalent. Across all surviving relationships, committed firms also accumulate, and the harmful pair's share can fall. Keeping the audit record separate from the event of survival is therefore necessary both for interpreting the learning mechanism and for understanding the selection it can generate.

\bmhead{Supplementary information}

Online Resource 1: \textit{Online Appendix to ``Audit Silence and the Capacity Trap.''} The appendix contains supplementary proofs, supporting lemmas, the exact two-period equilibrium calculations, and details of the computational verification.

\medskip
Online Resource 2: \textit{Replication code for ``Audit Silence and the Capacity Trap.} The archive contains a standalone Python script, execution instructions, and recorded results for the numerical examples and checks of filtering, clean-audit departures, patient-player bounds, capacity comparisons, welfare, and survivor selection.

\section*{Statements and Declarations}

\noindent\textbf{Funding.} This work was supported by the French National Research Agency (ANR) under grant ANR-17-EURE-0010 (Investissements d'Avenir program).

\medskip
\noindent\textbf{Competing interests.} The author declares no competing interests.

\medskip
\noindent\textbf{Author contributions.} Georgy Lukyanov is the sole author and is responsible for the research and writing.

\medskip
\noindent\textbf{Data availability.} No empirical datasets were generated or analyzed for this theoretical study.

\medskip
\noindent\textbf{Code availability.} The replication script and execution instructions are supplied in Online Resource 2. The script requires Python 3.9 or later and uses no external data or third-party packages.

\clearpage
\appendix
\setcounter{equation}{0}
\renewcommand{\theHequation}{appendix.\arabic{equation}}
\section*{Online Appendix}
\addcontentsline{toc}{section}{Online Appendix}
This appendix provides supporting arguments for ``Audit Silence and the Capacity Trap.'' Appendix A specifies histories and assessments and establishes existence of sequential equilibrium. Appendix B expands the filtering and incentive calculations for strategies that may depend on private action histories. Appendix C gives the terminal benchmark and an exact two-period equilibrium. Appendices D--F contain the continuation bounds, clean-audit calculations, and patient-player comparison. Appendices G and H derive the capacity, welfare, and survivor-selection results, making explicit the histories over which behavior is determined and the information retained by the observer. Appendix I records numerical examples and the scope of the verification.

\section{Histories, assessments, and existence}\label{app-app:equilibrium}

We begin by specifying the histories and assessments used in the infinite-horizon game. These definitions allow strategies to depend on each player's own past actions. We then establish that the set of sequential equilibria is nonempty. The entry theorem applies to every such equilibrium without selecting one, so existence is a necessary preliminary to its interpretation.

\subsection{Histories and strategies}

The persistent firm type is \(\tau_F\in\{c,s\}\), where \(c\) is committed to compliance and \(s\) is strategic. The persistent inspector type is \(\tau_I\in\{k,f\}\), where \(k\) is capacity constrained and \(f\) is a functioning strategic inspector. The initial distribution is
\[
\Prob(\tau_F=c)=\lambda_0,\qquad
\Prob(\tau_I=k)=\mu_0,
\]
and the two types are independent.

An active public history of length \(t\) is
\[
h_t=(y_0,\ldots,y_{t-1})
\in\{\No,\Ac\}^{t}.
\]
The signal \(\Det\) terminates the game and therefore has no active successor. The strategic firm's private history appends its own past actions to \(h_t\); the functioning inspector's private history appends its own past effort choices. There are no other private signals. In particular, the reduced-form flow-utility number is not separately observed before the next action.

A behavioral strategy of the strategic firm assigns a probability of \(\V\) at each of its information sets. A behavioral strategy of the functioning inspector assigns a probability of \(\Hi\) at each of its information sets. The committed firm chooses \(\C\) with probability one. The constrained inspector has no effort choice and implements an audit with probability \(q\).

The audit probability conditional on functioning effort \(e\) is
\[
\theta_I(e)=
\begin{cases}
\al,&e=\Lo,\\
\ah,&e=\Hi.
\end{cases}
\]
Conditional on constrained capacity it is \(\theta_I=q\). Given firm action \(z\in\{\C,\V\}\) and audit propensity \(\theta\), the public-signal probabilities are
\begin{align}
\Prob(\No\mid z,\theta)&=1-\theta,\label{app-eq:kernel-N}\\
\Prob(\Ac\mid z,\theta)&=\theta
  \{1-d\mathbf{1}_{\{z=\V\}}\},\label{app-eq:kernel-A}\\
\Prob(\Det\mid z,\theta)&=\theta d
  \mathbf{1}_{\{z=\V\}}.\label{app-eq:kernel-D}
\end{align}

An assessment consists of behavioral strategies and beliefs at every information set. Sequential equilibrium has its usual meaning: strategies are sequentially rational and beliefs are limits of Bayes-consistent beliefs under completely mixed perturbations.

\subsection{Non-vacuity of the infinite-horizon analysis}

The existence argument proceeds through finite truncations. Each truncation admits a sequential equilibrium, and compactness provides a pointwise limit of a suitable sequence of assessments. Two properties must be checked at that limit. Beliefs must remain consistent with completely mixed perturbations, and continuation strategies must remain optimal against arbitrary deviations. Discounting gives a uniform bound on distant payoffs, allowing the finite-game optimality inequalities to pass to the limit.

\begin{proposition}\label{app-prop:existence}
The infinite-horizon discounted game has a sequential equilibrium.
\end{proposition}

\begin{proof}
Stage payoffs are bounded, action and signal sets are finite, every player has perfect recall, and the set of finite histories is countable. Let \(\Gamma^n\) be the game truncated after period \(n\), with zero terminal continuation payoffs. By the finite-game existence theorem, \(\Gamma^n\) has a sequential equilibrium \((s^n,\eta^n)\).

Extend \(s^n\) beyond date \(n\) by a fixed completely mixed reference strategy and assign arbitrary beliefs after date \(n\) to the unperturbed assessment. At information sets through date \(n\), finite-game consistency provides completely mixed strategies whose Bayes beliefs converge to \(\eta^n\). Because there are only finitely many such information sets, choose one perturbation \(\widetilde s^n\) whose action probabilities and induced beliefs are within \(1/n\) of \(s^n\) and \(\eta^n\), respectively, at every information set through date \(n\); use the same completely mixed reference strategy thereafter.

The product of the behavioral-strategy simplexes over the countable collection of information sets is compact and metrizable under pointwise convergence. The same is true of the product of the finite-dimensional belief simplexes. A diagonal subsequence of the extended assessments \((s^n,\eta^n)\) therefore converges pointwise to an assessment \((s^*,\eta^*)\). On every fixed finite collection of information sets, \(\widetilde s^n\) and its Bayes beliefs have the same limit. Hence \((s^*,\eta^*)\) is consistent: it is the limit of assessments generated by completely mixed strategies on the infinite tree.

It remains to establish sequential rationality. Fix an information set \(I\) at date \(t\) and a deviation that differs from \(s^*\) for only finitely many subsequent periods. For all sufficiently large \(n\), the corresponding finite deviation is available in \(\Gamma^n\), so the sequential-rationality inequality holds under \((s^n,\eta^n)\). Expected discounted payoffs are continuous under pointwise convergence on every finite subtree. The remaining tail is bounded uniformly by
\[
\frac{2M\delta^{R+1}}{1-\delta}
\]
after \(R\) periods from \(I\), where \(M\) bounds absolute stage payoffs. First pass to the diagonal limit on the finite subtree and then let \(R\to\infty\). The equilibrium inequality therefore holds for every finite deviation at \(I\).

Finally, truncate any infinite continuation deviation after \(L\) periods and then return to \(s^*\). Relative to the original deviation, the payoff error is at most
\[
\frac{2M\delta^{L+1}}{1-\delta}.
\]
Letting \(L\to\infty\) proves sequential rationality against arbitrary deviations. Since \(I\) was arbitrary, \((s^*,\eta^*)\) is a sequential equilibrium.
\end{proof}

\section{Belief factorization and dynamic incentives}
\label{app-app:beliefs}

\subsection{Factorization with private action histories}

The main paper's filtering lemma permits arbitrary dependence on private action histories. We provide the expanded argument here. The induction keeps track of both persistent type and private action history for each player, since a player's future randomization may depend on either. Factorization at this level implies the required equality between public and private beliefs about the opponent.

\begin{lemma}\label{app-lem:product-filter}
At every active public history \(h_t\), the conditional joint distribution of
\[
(\tau_F,\text{firm private history})
\quad\text{and}\quad
(\tau_I,\text{inspector private history})
\]
factors into its two marginals. Consequently, each player's posterior about the opponent's persistent type is independent of its own private action history and equals the corresponding public marginal posterior.
\end{lemma}

\begin{proof}
The claim holds at \(t=0\) by independent types. Suppose it holds at an active public history \(h_t\). Conditional on the two private states and \(h_t\), the players' behavioral randomizations are independent. For a realized pair of current actions, the likelihood of each nonterminal public signal factors into a firm term and an inspector term:
\[
L_{\No}(z,\theta)
=1\cdot(1-\theta),
\]
and
\[
L_{\Ac}(z,\theta)
=\{1-d\mathbf{1}_{\{z=\V\}}\}\cdot\theta.
\]
The action probabilities themselves also factor by player. Therefore the unnormalized conditional mass of the two augmented private histories after either \(\No\) or \(\Ac\) is a product of a firm-side mass and an inspector-side mass. Normalization preserves the product. Induction proves the claim after every active public history. Marginalizing private histories gives independence of persistent types. Conditioning additionally on one's own private history leaves the opponent marginal unchanged.

The conclusion also applies at a zero-probability private history in a sequential equilibrium. For every completely mixed perturbation used to establish consistency, the product identity holds exactly and the opponent marginal is independent of that private history. Passing to the perturbation limit preserves the identity.
\end{proof}

\begin{lemma}
\label{app-lem:private-value}
Fix an active public history in a sequential equilibrium. A strategic player's continuation value is the same at all of its private action histories consistent with that public history.
\end{lemma}

\begin{proof}
Consider two private histories of one player that share the public history. By Lemma \ref{app-lem:product-filter}, the conditional distribution of the opponent's persistent type and private history is identical at the two information sets. Own past actions are sunk, do not affect any future payoff or signal kernel, and are not observed by the opponent. The feasible continuation plans from the two information sets are therefore in a payoff-preserving bijection. Against the opponent's fixed continuation strategy, the two best-response problems have the same value. Sequential rationality requires the equilibrium continuation plan to attain that value at each information set. The continuation values are consequently equal, even if different private histories select different optimal actions.
\end{proof}

\begin{remark}
The terminal signal also has product likelihood
\[
L_{\Det}(z,\theta)
=d\mathbf{1}_{\{z=\V\}}\cdot\theta.
\]
Nothing in the filtering argument fails at detection, but no continuation belief is needed because the relationship ends.
\end{remark}

\subsection{Posterior formulas under history-averaged strategies}

At an active public history, use action probabilities averaged over private histories and write
\[
a=(1-\lambda)\sigma,
\qquad
x=\al+(\ah-\al)\rho,
\qquad
p=\mu q+(1-\mu)x.
\]
Conditional on an audit, the probability of no finding is \(1-da\). Bayes' rule gives
\begin{align*}
\Prob(\tau_F=c\mid\Ac)
&=\frac{\lambda}{1-da},\\
\Prob(\tau_I=k\mid\Ac)
&=\frac{\mu q(1-da)}{p(1-da)}
=\frac{\mu q}{p}.
\end{align*}
The likelihood of no audit is independent of the firm's conduct. Its firm-side factor is therefore one, and the compliance belief remains unchanged:
\begin{align*}
\Prob(\tau_F=c\mid\No)&=\lambda,\\
\Prob(\tau_I=k\mid\No)
&=\frac{\mu(1-q)}
{\mu(1-q)+(1-\mu)(1-x)}
=\frac{\mu(1-q)}{1-p}.
\end{align*}
These are the posterior formulas stated in the belief lemma of the main text.

Full support follows from
\[
d<1,\qquad 0<q,\al,\ah<1.
\]
For every active type pair and every action pair, \(\No\) occurs with positive probability. The signal \(\Ac\) also occurs with positive probability because an audit can be implemented and detection can fail.

\subsection{Deviation gains}

\begin{lemma}\label{app-lem:dynamic-diffs}
At any strategic-firm information set, let \(p\) be its conditional audit probability. Its violation gain is
\[
\Delta_F
=g-pd(g+f)-\delta pdV_{\Ac}.
\]
At any functioning-inspector information set, let \(a\) be its conditional aggregate violation probability. The sign of its high-effort gain is the sign of
\[
\Psi
=-c+ad(r+\ell)
+\delta\{(1-da)W_{\Ac}-W_{\No}\}.
\]
\end{lemma}

\begin{proof}
Lemma \ref{app-lem:private-value} implies that continuation after a given public signal is independent of the player's own current private action. Thus compliance gives
\[
Q_{\C}
=\delta\{pV_{\Ac}+(1-p)V_{\No}\}.
\]
Violation gives
\[
Q_{\V}
=(1-p)(g+\delta V_{\No})
+p\{(1-d)(g+\delta V_{\Ac})-df\}.
\]
Subtracting cancels the no-audit continuation and yields the first expression.

Conditional on an implemented audit, the functioning inspector's expected payoff is
\[
B_A
=-c+a\{dr-(1-d)\ell\}
+\delta(1-da)W_{\Ac}.
\]
Conditional on no audit, it is
\[
B_N=-a\ell+\delta W_{\No}.
\]
High rather than low effort raises the audit probability by \(\ah-\al>0\), so its payoff gain is \((\ah-\al)(B_A-B_N)\). Simplification gives the displayed \(\Psi\).
\end{proof}

\section{Terminal play and the exact two-period equilibrium}\label{app-app:two-period}

\subsection{The one-period benchmark}

Define
\begin{equation}\label{app-eq:app-static-thresholds}
p^*=\frac{g}{d(g+f)},\qquad
a^*=\frac{c}{d(r+\ell)}.
\end{equation}

\begin{proposition}\label{app-prop:static}
In the final period, the strategic firm strictly prefers \(\V\) when \(p<p^*\) and strictly prefers \(\C\) when \(p>p^*\). The functioning inspector strictly prefers \(\Hi\) when \(a>a^*\) and strictly prefers \(\Lo\) when \(a<a^*\).

An interior terminal equilibrium exists at a state \((\lambda,\mu)\) if
\[
a^*<1-\lambda
\]
and
\[
\mu q+(1-\mu)\al<p^*
<\mu q+(1-\mu)\ah.
\]
It has aggregate actions \(p=p^*\) and \(a=a^*\).
\end{proposition}

\begin{align}
\sigma^*&=\frac{a^*}{1-\lambda},\label{app-eq:sigma-terminal}\\
\rho^*&=
\frac{(p^*-\mu q)/(1-\mu)-\al}{\ah-\al}.
\label{app-eq:rho-terminal}
\end{align}
The strategic firm's terminal value is zero and the functioning inspector's terminal value is \(-\ell a^*\).

\begin{proof}
Set \(\delta=0\) in Lemma \ref{app-lem:dynamic-diffs}. This yields the two thresholds and the mixing probabilities. At \(p=p^*\), both firm actions give zero. At \(a=a^*\), the inspector is indifferent between an audit realization and no audit:
\[
-c+a^*\{dr-(1-d)\ell\}
=-a^*\ell.
\]
Every effort mixture therefore gives value \(-\ell a^*\).
\end{proof}

The terminal thresholds illustrate how low expected enforcement can coexist with a strong incentive to audit. If the capacity posterior is high enough that even maximum functioning effort gives \(\mu q+(1-\mu)\ah<p^*\), the strategic firm strictly prefers to violate. When the resulting violation probability exceeds \(a^*\), the functioning inspector strictly prefers high effort. The audit rate nevertheless remains below deterrence because the firm places substantial probability on constrained capacity. The infinite-horizon argument must additionally account for continuation incentives and explain how beliefs reach the required range.

\subsection{An exact two-period capacity trap}\label{app-sec:two-period}

This section constructs a particular equilibrium transition in a two-period game. It illustrates how a missed audit and a clean audit can lead to different continuation behavior. It is distinct from the main paper's entry theorems, which give a sufficient history for entry across all equilibria under additional restrictions. There are two decision periods, detection in either one ends the relationship, and otherwise the second period is terminal. Both strategic players mix in the first period; a single silent observation then puts the continuation into the trap, while a clean audit puts it into an interior equilibrium instead.

Assume
\begin{equation}\label{app-eq:two-order}
q,\al<p^*<\ah.
\end{equation}
Define the terminal capacity threshold
\begin{equation}\label{app-eq:muc}
\mu^c=\frac{\ah-p^*}{\ah-q}
\end{equation}
and the no-audit Bayes factor
\[
b=\frac{1-p^*}{1-q}.
\]
Thus \(\mu>\mu^c\) exactly when maximum functioning effort cannot raise the aggregate audit rate to \(p^*\).

At a terminal capacity-trap state in which both strategic players use their aggressive actions, the functioning inspector's value is
\begin{equation}\label{app-eq:wcap}
W^{\mathrm{cap}}
=-\ah c+(1-\lambda)\{\ah d(r+\ell)-\ell\}.
\end{equation}
Finally define
\begin{equation}\label{app-eq:ahat}
\widehat a
=\frac{c+\delta\ell a^*+\delta W^{\mathrm{cap}}}
{d\{r+\ell+\delta\ell a^*\}}.
\end{equation}

\begin{proposition}\label{app-prop:two-period}
Suppose
\begin{align}
b\mu^c&<\mu<\mu^c,\label{app-eq:two-cond1}\\
1-\lambda&>a^*,\qquad
0<\widehat a<1-\lambda,\label{app-eq:two-cond2}\\
1-\frac{\lambda}{1-d\widehat a}&>a^*.\label{app-eq:two-cond3}
\end{align}
There is a two-period sequential equilibrium with the following behavior.

In the first period both strategic players mix, with
\[
p=p^*,\qquad a=\widehat a,
\]
\[
\sigma=\frac{\widehat a}{1-\lambda},\qquad
\rho=
\frac{(p^*-\mu q)/(1-\mu)-\al}{\ah-\al}.
\]
After a clean audit, the terminal equilibrium is interior:
\[
p_{\Ac}=p^*,\qquad a_{\Ac}=a^*.
\]
After no audit,
\[
\mu_{\No}=\frac{\mu(1-q)}{1-p^*}=\frac{\mu}{b}>\mu^c.
\]
Both strategic players then choose their aggressive actions,
\[
\sigma_{\No}=\rho_{\No}=1,
\]
but expected enforcement remains below the deterrence threshold:
\[
p_{\No}=\mu_{\No}q+(1-\mu_{\No})\ah<p^*,
\qquad a_{\No}=1-\lambda>a^*.
\]
The probability of a terminal undetected violation is strictly greater after no audit than after a clean audit.
\end{proposition}

The conditions ensure that the proposed mixing probabilities and terminal responses are feasible. Condition \eqref{app-eq:two-cond1} places the initial capacity belief below the terminal threshold but close enough that one missed audit takes it above that threshold. The initial audit rate \(p^*\) is then attainable, while maximum functioning effort after silence cannot attain it. A clean audit moves the capacity belief in the other direction. Condition \eqref{app-eq:two-cond3} ensures that, after the accompanying increase in the compliance belief, enough strategic-firm probability remains to support the terminal interior equilibrium.

The initial indifference conditions use the values of these two continuations. After a clean audit the strategic firm's terminal value is zero, so initial firm indifference requires \(p=p^*\). The inspector's value after that signal is \(-\ell a^*\), whereas its value after no audit is \(W^{\mathrm{cap}}\). Substituting them into the inspector's deviation gain gives \eqref{app-eq:ahat}. The remaining strict inequalities keep the initial mixing probabilities interior.\footnote{The strict inequalities also ensure that the construction persists under sufficiently small perturbations of the implementation and detection parameters \((d,q,\al,\ah)\). Appendix~\ref{app-app:verification} records the scope of this robustness statement.}

\begin{example}\label{app-ex:two-period}
Let
\[
(g,f,c,r,\ell,\delta,d)=(1,4,0.3,1,1,0.9,0.98)
\]
and
\[
(q,\al,\ah,\lambda,\mu)=(0.02,0.10,0.98,0.05,0.72).
\]
All audit probabilities have full support and capacity separation holds. The equilibrium is:
\begin{center}
\centering
\begin{tabular}{@{}lrrrr@{}}
\toprule
History & \(p\) & \(a\) & \(\rho\) & \(\sigma\)\\
\midrule
Initial & 0.20408 & 0.45844 & 0.65618 & 0.48257\\
Clean audit & 0.20408 & 0.15306 & 0.13416 & 0.16835\\
No audit & 0.12894 & 0.95000 & 1 & 1\\
\bottomrule
\end{tabular}
\end{center}
The undetected-violation probability is \(0.12245\) after a clean audit and \(0.82996\) after no audit.
\end{example}

\subsection{Proof of Proposition~\ref{app-prop:two-period}}

\begin{proof}
\textit{Step 1: first-period feasibility.} The aggregate audit rate under functioning low effort is below \(p^*\), because both \(q\) and \(\al\) are below \(p^*\). Moreover,
\[
\mu<\mu^c
\quad\Longleftrightarrow\quad
\mu q+(1-\mu)\ah>p^*.
\]
Thus a strict functioning-inspector mixture can generate \(p^*\). The condition \(0<\widehat a<1-\lambda\) gives a strict firm mixture.

\textit{Step 2: no-audit successor.} If the first-period aggregate audit rate is \(p^*\), then
\[
\mu_{\No}
=\frac{\mu(1-q)}{1-p^*}
=\frac{\mu}{b}.
\]
The inequality \(\mu>b\mu^c\) implies \(\mu_{\No}>\mu^c\), hence
\[
p_{\No}^{\max}
=\mu_{\No}q+(1-\mu_{\No})\ah<p^*.
\]
The terminal firm therefore strictly chooses \(\V\), regardless of the functioning inspector's action. Aggregate violation is \(1-\lambda>a^*\), so the functioning inspector strictly chooses \(\Hi\). Its terminal value is
\[
W_{\No}
=-\ah c+(1-\lambda)\{\ah d(r+\ell)-\ell\}
=W^{\mathrm{cap}}.
\]

\textit{Step 3: clean-audit successor.} The posteriors are
\[
\lambda_{\Ac}
=\frac{\lambda}{1-d\widehat a},
\qquad
\mu_{\Ac}
=\frac{\mu q}{p^*}.
\]
Because \(q<p^*\), \(\mu_{\Ac}<\mu<\mu^c\). The lower aggregate audit bound remains below \(p^*\), while the upper bound is above it. The assumed inequality
\[
1-\lambda_{\Ac}>a^*
\]
supplies enough strategic-firm mass for interior mixing. Proposition~\ref{app-prop:static} therefore gives the terminal interior equilibrium
\[
p_{\Ac}=p^*,\qquad a_{\Ac}=a^*,
\]
with
\[
V_{\Ac}=0,\qquad W_{\Ac}=-\ell a^*.
\]

\textit{Step 4: first-period indifference.} Substituting \(V_{\Ac}=0\) into the firm difference gives
\[
\Delta_F=g-pd(g+f).
\]
Thus \(p=p^*\) makes the strategic firm indifferent.

Substitute the two inspector continuation values into its difference:
\[
\Psi
=-c+ad(r+\ell)
+\delta\{-(1-da)\ell a^*-W^{\mathrm{cap}}\}.
\]
Setting \(\Psi=0\) and solving for \(a\) gives
\[
a
=\frac{c+\delta\ell a^*+\delta W^{\mathrm{cap}}}
{d\{r+\ell+\delta\ell a^*\}}
=\widehat a.
\]
The strict feasibility inequalities established in Step 1 produce the stated \(\sigma\) and \(\rho\) in \((0,1)\).

\textit{Step 5: beliefs and sequential rationality.} All nonterminal histories have positive probability, so the stated beliefs follow from Bayes' rule. Steps 2--4 verify optimality at every active information set. At terminal detection there is no action. The assessment is therefore a sequential equilibrium.
\end{proof}

\begin{corollary}\label{app-cor:exact-risk}
In the equilibrium of Proposition \ref{app-prop:two-period},
\[
(1-\lambda)(1-dp_{\No})
>a^*(1-dp^*).
\]
\end{corollary}

\begin{proof}
The no-audit successor has
\[
1-\lambda>a^*
\quad\text{and}\quad
p_{\No}<p^*.
\]
Both positive factors in the left-hand product are therefore strictly larger than their counterparts on the right.
\end{proof}

\section{The global audit-silence theorem}\label{app-app:global}

The proof has three components. First, continuation-value bounds give sufficient conditions for violation and high effort at any information set. Second, the likelihood ratio after a missed audit has a lower bound that is independent of the functioning inspector's strategy. Third, repeated application of that bound takes the capacity posterior into the range needed for both incentive conditions, while the compliance belief stays unchanged. The following lemmas make each part explicit.

\subsection{Continuation-independent dominance}

\begin{lemma}\label{app-lem:firm-dom}
At every strategic-firm information set,
\[
0\leq V\leq\frac{g}{1-\delta}.
\]
If its perceived current audit probability satisfies
\[
p<\pi_\delta
\equiv
\frac{g(1-\delta)}
{d\{g+(1-\delta)f\}},
\]
then \(\V\) is strictly optimal, independently of continuation play.
\end{lemma}

\begin{proof}
Compliance forever guarantees zero. No active-period payoff exceeds \(g\), so the upper bound follows. At an arbitrary information set, the same one-shot deviation calculation as in Lemma \ref{app-lem:dynamic-diffs} applies because Lemma \ref{app-lem:private-value} removes dependence on the firm's own current action. Hence
\begin{align*}
\Delta_F
&=g-pd(g+f)-\delta pdV_{\Ac}\\
&\geq
g-pd(g+f)-\frac{\delta pdg}{1-\delta}.
\end{align*}
The last expression is positive exactly when \(p<\pi_\delta\).
\end{proof}

\begin{lemma}\label{app-lem:inspector-dom}
Let
\[
\kappa_L
=\al c+\max\{0,\ell-\al d(r+\ell)\},
\qquad
R=\max\{r-c,0\}.
\]
At every functioning-inspector information set,
\[
-\frac{\kappa_L}{1-\delta}\leq W\leq R.
\]
If its perceived aggregate violation probability satisfies
\[
a>\beta_\delta
\equiv
\frac{c+\delta\{\kappa_L/(1-\delta)+R\}}
{d(r+\ell)},
\]
then \(\Hi\) is strictly optimal, independently of continuation play.
\end{lemma}

\begin{proof}
If the functioning inspector chooses low effort, its expected current payoff at aggregate violation rate \(a\in[0,1]\) is
\[
-\al c+a\{\al d(r+\ell)-\ell\}
\geq-\kappa_L.
\]
Choosing low effort forever therefore guarantees \(W\geq-\kappa_L/(1-\delta)\).

All nonterminal inspector payoffs are nonpositive. The only possibly positive payoff is \(r-c\) upon detection, and detection terminates the game. Thus at most one positive payoff can occur and \(W\leq R\). It follows that
\[
(1-da)W_{\Ac}-W_{\No}
\geq-\frac{(1-da)\kappa_L}{1-\delta}-R
\geq-\frac{\kappa_L}{1-\delta}-R.
\]
Consequently,
\[
\Psi
\geq
-c+ad(r+\ell)
-\delta\left\{\frac{\kappa_L}{1-\delta}+R\right\}.
\]
The right-hand side is positive when \(a>\beta_\delta\).
\end{proof}

\subsection{Likelihood-ratio drift under arbitrary strategies}

Assume
\[
0<q<\al<\ah<1,\qquad
q<\pi_\delta<\ah,\qquad
\beta_\delta<1.
\]
Define
\[
\mu^\dagger=\frac{\ah-\pi_\delta}{\ah-q},
\quad
O^\dagger=\frac{\mu^\dagger}{1-\mu^\dagger},
\quad
F=\frac{1-q}{1-\al}>1.
\]

\begin{lemma}\label{app-lem:odds}
At any active public history, let \(x_t\) be the functioning inspector's current audit probability after averaging over its private histories. Then
\[
x_t\in[\al,\ah].
\]
After no audit, the public odds of constrained capacity satisfy
\[
O_{t+1}=O_t\frac{1-q}{1-x_t}\geq O_tF.
\]
Along a run of no audits, the firm-type posterior \(\lambda\) is unchanged.
\end{lemma}

\begin{proof}
Every pure functioning effort produces an audit probability in \(\{\al,\ah\}\); arbitrary mixing and averaging over private histories stays in their convex hull. Under constrained capacity, the probability of no audit is \(1-q\). Under the functioning type it is \(1-x_t\). Bayes' rule gives the odds equality, and \(x_t\geq\al\) gives the inequality. Product-form filtering in Lemma \ref{app-lem:product-filter} implies that no audit has a constant firm-side likelihood, so it does not update \(\lambda\).
\end{proof}

\begin{theorem}\label{app-thm:global-app}
Suppose the current public beliefs satisfy
\[
\mu_0\in(0,1),\qquad
\lambda_0<1-\beta_\delta.
\]
Let
\[
O_0=\frac{\mu_0}{1-\mu_0},
\qquad
K=\min\{k\geq0:O_0F^k>O^\dagger\}.
\]
After \(K\) consecutive no-audit observations, every sequential equilibrium prescribes \(\V\) at every strategic-firm information set and \(\Hi\) at every functioning-inspector information set. Every further no-audit observation preserves this conclusion.
\end{theorem}

\begin{proof}
Lemma \ref{app-lem:odds} gives
\[
O_K\geq O_0F^K>O^\dagger,
\]
so \(\mu_K>\mu^\dagger\). At every strategic-firm information set consistent with the public history, Lemma \ref{app-lem:product-filter} gives the same posterior \(\mu_K\) about inspector type. Even if the functioning type chooses high effort, the firm's audit probability is bounded above by
\[
\mu_Kq+(1-\mu_K)\ah<\pi_\delta.
\]
Lemma \ref{app-lem:firm-dom} makes \(\V\) strictly optimal at every such information set.

Since every strategic firm violates, aggregate violation from any inspector information set is
\[
a=1-\lambda_K=1-\lambda_0>\beta_\delta.
\]
The equality of private and public opponent-type beliefs again follows from Lemma \ref{app-lem:product-filter}. Lemma \ref{app-lem:inspector-dom} makes \(\Hi\) strictly optimal at every functioning-inspector information set.

Under high effort, one more no-audit signal multiplies constrained-capacity odds by
\[
\frac{1-q}{1-\ah}>1.
\]
The posterior \(\mu\) rises, \(\lambda\) remains fixed, and both strict dominance inequalities continue to hold. Induction proves invariance after further silence.
\end{proof}

\begin{remark}
The constrained inspector alone generates \(K\) no-audit observations with probability \((1-q)^K\). Hence, at the initial public state,
\[
\Prob(\No^K)\geq\mu_0(1-q)^K>0.
\]
The joint event that the inspector is constrained, the firm is strategic, and the next \(K\) observations are all no audits has probability
\[
\mu_0(1-\lambda_0)(1-q)^K.
\]
\end{remark}

\begin{remark}
If \(\al<q\), then when the functioning inspector uses low effort,
\[
\frac{1-q}{1-\al}<1.
\]
A no-audit observation lowers rather than raises the odds of constrained capacity. If \(\al=q\), the signal is uninformative under low effort. Full support alone therefore cannot produce a uniform finite-entry bound.
\end{remark}

\subsection{Hazard and one-period welfare}

Inside the dominance region,
\[
p(\mu)=\mu q+(1-\mu)\ah,\qquad a=1-\lambda.
\]
Following no audit,
\[
\frac{\mu'}{1-\mu'}
=\frac{\mu}{1-\mu}\frac{1-q}{1-\ah}.
\]
Since \(q<\ah\), \(\mu'>\mu\), and therefore \(p(\mu')<p(\mu)\). The detection hazard
\[
\chi(\mu)=da\,p(\mu)
\]
strictly falls, while undetected violation
\[
u(\mu)=a\{1-dp(\mu)\}
\]
strictly rises.

For the social loss
\[
\mathcal L(p)
=\mathcal H a(1-dp)+C_sp+K_0dap,
\]
direct subtraction gives
\[
\mathcal L(p')-\mathcal L(p)
=(p-p')\{da(\mathcal H-K_0)-C_s\}.
\]
This is positive when
\[
da(\mathcal H-K_0)>C_s.
\]

\section{Clean-audit successors and departure}\label{app-app:clean-exit}

A missed audit leaves the firm's reputation unchanged. A clean audit provides evidence about both parties, and that difference determines whether the preceding result can continue to apply. Starting inside \(\Tset\), write \(p=p^H(\mu)\). Since all strategic firms violate, the successor beliefs are
\begin{equation}\label{app-eq:clean-in-trap}
\lambda_A=\frac{\lambda}{1-d(1-\lambda)},\qquad
\mu_A=\frac{\mu q}{\mu q+(1-\mu)\ah}.
\end{equation}
The first posterior rises and the second falls. A clean audit suggests that the firm may be committed to compliance and that the inspector may possess functioning capacity. Both changes weaken the sufficient conditions for the configuration of universal strategic violation and high functioning effort.

\begin{proposition}\label{app-prop:clean-exit}
Under the hypotheses of Theorem~\ref{app-thm:global-app}, start at a history in \(\Tset\). After a clean audit the successor remains in \(\Tset\) if and only if
\begin{align}
\lambda&<\lambda^A_\delta
\equiv\frac{(1-\beta_\delta)(1-d)}{1-d+d\beta_\delta},\label{app-eq:lambda-clean-cutoff}\\
\mu&>\mu^A_\delta
\equiv\frac{\mu^\dagger\ah}
{q(1-\mu^\dagger)+\mu^\dagger\ah}.\label{app-eq:mu-clean-cutoff}
\end{align}
The new cutoffs satisfy \(0<\lambda^A_\delta<1-\beta_\delta\) and \(\mu^\dagger<\mu^A_\delta<1\). If either strict inequality fails, the successor is outside the sufficient region, including when equality holds.
\end{proposition}

\begin{proof}
Apply the posterior formulas established above with \(a=1-\lambda\) and \(x=\ah\), obtaining \eqref{app-eq:clean-in-trap}. All denominators are positive. Rearranging \(\lambda_A<1-\beta_\delta\) gives
\[
\lambda\{1-d+d\beta_\delta\}<(1-\beta_\delta)(1-d).
\]
Rearranging \(\mu_A>\mu^\dagger\) gives
\[
\mu\{q(1-\mu^\dagger)+\mu^\dagger\ah\}>\mu^\dagger\ah.
\]
These are exactly the two displayed conditions. The cutoff orderings follow from \(d,\beta_\delta,\mu^\dagger\in(0,1)\) and \(q<\ah\).
\end{proof}

The proposition identifies departure from a set of sufficient conditions. It does not imply that compliance or low effort is optimal immediately after departure. Those choices depend on continuation values that the bounds do not determine. The distinction is important: the posterior calculation is exact, while the region itself is a sufficient characterization.

\begin{corollary}\label{app-cor:clean-reentry}
If a clean audit from \(\Tset\) yields \(\lambda_A<1-\beta_\delta\) but \(\mu_A\leq\mu^\dagger\), a sufficiently long subsequent uninterrupted run of missed audits returns beliefs to \(\Tset\) in every sequential equilibrium. The return run has positive conditional probability, with its sufficient length obtained from the finite-entry formula above using \(\mu_A\).

If instead \(\lambda_A\geq1-\beta_\delta\), no later active public history can belong to this same sufficient region.
\end{corollary}

\begin{proof}
The first case satisfies the compliance-belief hypothesis of Theorem~\ref{app-thm:global-app} at the clean-audit successor. In the second case, the posterior formulas imply that every subsequent missed audit leaves \(\lambda\) unchanged and every clean audit weakly increases it. Thus the strict compliance-belief condition can never be restored.
\end{proof}

There are therefore two distinct reasons for departure. Confidence in functioning capacity can recover after an audit and subsequently fall again after silence. Confidence in committed compliance, once sufficiently high, removes the hypothesis needed for universal high effort in this particular dominance argument. The latter is a permanent loss of this sufficient guarantee along active histories; it is not a claim that capacity-trap behavior can never occur again outside the certified region.

\begin{proposition}\label{app-prop:finite-clean-exit}
The sufficient region \(\Tset\) is not absorbing. From any history in it, there is a finite sequence of clean audits, with positive conditional probability, whose last observation first takes beliefs outside \(\Tset\).
\end{proposition}

\begin{proof}
Write the starting compliance and capacity odds as \(L=\lambda/(1-\lambda)\) and \(O=\mu/(1-\mu)\). As long as each preceding history remains in \(\Tset\), a clean audit multiplies these odds by \((1-d)^{-1}\) and \(q/\ah\), respectively. Hence after \(n\) consecutive clean audits, up to and including the first departure,
\[
L_n=L(1-d)^{-n},\qquad O_n=O(q/\ah)^n.
\]
The first departure occurs at
\[
m=\min\left\{n\geq1:
L(1-d)^{-n}\geq\frac{1-\beta_\delta}{\beta_\delta}
\ \text{or}\ O(q/\ah)^n\leq O^\dagger\right\}.
\]
This integer is finite, since \(L>0\), \(d>0\), and \(q/\ah<1\). Before that observation both actions are fixed by dominance. The probability of the specified \(m\) clean audits is therefore
\[
\{\lambda+(1-\lambda)(1-d)^m\}
\{\mu q^m+(1-\mu)\ah^m\}>0.
\]
The product follows by conditioning on the initially independent types at the starting public history. A clean audit is nonterminal, so the last observation is an active departure rather than termination.
\end{proof}

One clean audit can leave a sufficiently deep history inside the region. A finite run of clean audits nevertheless produces departure. Conversely, any finite run consisting only of missed audits preserves membership once it is reached. These observations explain the sense in which the configuration is a trap sustained by silence without being an absorbing state of the game.

\section{The patient-player extension}\label{app-app:patient}

The patient-player argument uses two features of the model beyond the simple payoff bounds. A positive audit floor limits the strategic firm's value from continued violation. Once the capacity belief makes violation unavoidable over a sufficiently long finite interval, the inspector's incentives over that interval can be compared with a control problem in which firm behavior is fixed. We first establish the two supporting bounds and then justify the comparison for arbitrary inspector continuation plans.

\subsection{The firm bound and the auxiliary control problem}

\begin{proof}[Proof of the firm bound]
Every inspector type and effort choice produces an audit with probability at least \(q\). Given continuation values bounded above by \(\overline V_\delta\), compliance yields at most \(\delta\overline V_\delta\), whereas violation yields at most
\[
g-qd(g+f)+\delta(1-qd)\overline V_\delta
=\overline V_\delta.
\]
The numerator in main-text equation~\eqref{eq:firm-floor-value} is positive. Finite-horizon induction and the bounded discounted tail establish the upper bound for every strategy; compliance forever supplies the lower bound. Substitution in main-text equation~\eqref{eq:firm-diff} gives main-text equation~\eqref{eq:patient-pi}. Finally,
\[
\widehat\pi_\delta-q
=\frac{(1-\delta)\{g-qd(g+f)\}}{d\{g+(1-\delta)f\}}>0.
\]
The other strict inequality follows from \(\overline V_\delta>0\) and \(\delta>0\).
\end{proof}

In the auxiliary problem, write \(a=1-\lambda\) for the probability of violation and use the low-effort values \(w_C^L,w_V^L\) defined in the main paper. Each firm type follows a prescribed action, so the inspector does not need to predict a further strategic response by the firm. The compliance belief \(\lambda\) is therefore sufficient for this control problem. Its Bellman operator is a contraction on bounded functions because \(\delta<1\), and its finite action set guarantees a maximizer. After no audit the belief is unchanged, which allows the Bellman equation to be rearranged as below. After a clean audit the belief changes, and the inspector remains free to choose a different effort level.

\begin{proof}[Proof of the auxiliary inspection lemma]
Write \(\lambda'=\lambda/(1-da)\), and let
\[
B=-c+ad(r+\ell)+\delta(1-da)W^\circ(\lambda').
\]
The Bellman equation, using the self-loop after a missed audit, is equivalent to
\begin{equation}\label{app-eq:aux-Bellman}
W^\circ(\lambda)=
\max_{x\in\{\al,\ah\}}
\frac{-a\ell+xB}{1-\delta+\delta x}.
\end{equation}
The difference between the two fractions has the sign of \(S=(1-\delta)B+\delta a\ell\).

Let \(W_L(\lambda)=\lambda w_C^L+(1-\lambda)w_V^L\). If all future effort is low, its one-period effort gain is
\[
-c+ad(r+\ell)+\delta\{(1-da)W_L(\lambda')-W_L(\lambda)\}
=-c+aT_\delta.
\]
If \(a\leq a_\delta^\circ\), all subsequent active beliefs also satisfy this inequality, since clean audits only reduce \(a\). Thus low effort verifies the Bellman equation throughout the reachable region, proving \(W^\circ=W_L\) there. It is uniquely optimal when the inequality is strict.

If \(a>a_\delta^\circ\), feasibility of always low effort gives \(W^\circ(\lambda')\geq W_L(\lambda')\). Substituting this lower bound into \(S\) yields
\[
S\geq(1-\delta+\delta\al)(-c+aT_\delta)>0.
\]
High effort therefore uniquely maximizes \eqref{app-eq:aux-Bellman}. Its effort gain is \(B-\delta W^\circ(\lambda)=S/(1-\delta+\delta\ah)\), establishing main-text equation~\eqref{eq:aux-margin}. Finally, the denominator minus the numerator in main-text equation~\eqref{eq:aux-cutoff} is \((1-\delta)(dr-c)+d\ell\). It is positive at both endpoints \(\delta=0,1\) under main-text equation~\eqref{eq:patient-primitives}, and hence throughout the interval.
\end{proof}

\subsection{Comparison under arbitrary continuation strategies}

Fix an equilibrium assessment and an active starting history with capacity odds larger than \(O_I=O_F(\ah/q)^{M+1}\). Public posteriors are those of this fixed assessment. After every missed audit the odds multiplier is at least \((1-q)/(1-\al)>1\); after every clean audit it is at least \(q/\ah\). Thus every active public history at depth \(j\leq M+1\) has odds strictly above \(O_F\). The firm bound forces strategic violation at every information set at each of these histories. This is a conclusion about all the histories in a finite tree, rather than only the realized no-audit path.

Now consider an inspector information set immediately after either initial nonterminal signal. The opponent's type distribution depends only on that successor's public history, by the private-history factorization result. Its past private actions do not affect current conduct in the finite tree: committed firms comply and every strategic firm violates. From this inspector information set, the next \(M\) decision periods therefore have exactly the same controlled signal and payoff distributions as the auxiliary problem initialized at that successor's compliance belief.

The comparison must hold for alternative inspector plans as well as for its equilibrium plan. When the inspector deviates, the opponent's equilibrium strategy and the assessment's public updating functions are held fixed. The deviation may change which public history occurs, but every active history within the specified depth still makes strategic violation strictly optimal. Moreover, the private-history filtering result supplies the same opponent distribution at each relevant information set, including under consistent limiting beliefs at private histories of zero probability. The auxiliary problem permits the same continuation randomizations. It can therefore represent every inspector plan needed for the payoff comparison.

For completeness, let \(J\) be the actual optimal continuation value and \(J^\circ\) the auxiliary optimal value at this successor. The payoff under any plan lies in
\[
[L_\delta,R],\qquad L_\delta=-\frac{\ell+c}{1-\delta},
\quad R=\max\{r-c,0\}.
\]
The lower bound follows from the worst possible negative stage payoff. For the upper bound, every nonterminal payoff is nonpositive and the sole potentially positive terminal payoff can occur only once. Apply a given feasible effort plan in both problems for the first \(M\) periods, using the same action, implementation, and detection draws. Their payoffs and stopping events agree through these periods. If the relationship remains active, the two remaining values differ by at most \(R-L_\delta=C_\delta\). Taking expectations gives a difference of at most \(\delta^M C_\delta\). Matching a plan in each direction and taking suprema proves
\[
|J-J^\circ|\leq\delta^M C_\delta.
\]
The inequality is obtained by matching each feasible plan in one problem with a plan in the other and then taking the best attainable payoff on each side. It therefore compares optimal continuation values, including plans that depend on private histories. The use of a one-dimensional belief in the auxiliary control problem does not restrict equilibrium strategies in the original game to be Markov.

Apply this inequality at both one-period successors. With \(a=1-\lambda\), the original inspector's gain per unit audit probability differs from its auxiliary gain by at most
\[
\delta\bigl[(1-da)|W_A-W_A^\circ|+|W_N-W_N^\circ|\bigr]
\leq2C_\delta\delta^{M+1}.
\]
This is strictly smaller than \(\eta_\delta\), which proves high-effort dominance. The actual payoff difference between high and low effort multiplies both sides by the positive constant \(\ah-\al\).

\subsection{Clean audits in the patient-player region}

For each fixed choice of \(\Lambda\) and \(M\), write \(\mu_I=O_I/(1+O_I)\). The patient-player region is a rectangle, \(\lambda<\Lambda\) and \(\mu>\mu_I\), on which the same behavior as in \(\Tset\) is forced. After a clean audit it remains in this rectangle precisely when
\[
\lambda<\frac{\Lambda(1-d)}{1-d\Lambda},
\qquad
\mu>\frac{\mu_I\ah}{q(1-\mu_I)+\mu_I\ah}.
\]
These conditions follow by rearranging the same posterior formulas as in the main-text clean-audit proposition. If departure is solely through capacity beliefs, later silence restores membership with the bound using \(O_I\) and the updated initial odds. If the successor has \(\lambda_A\geq\Lambda\), this particular rectangle cannot be reached again because the compliance belief never falls on active histories. A larger cap may remain admissible when \(\lambda_A<1-a_\delta^\circ\); leaving a chosen rectangle is therefore not the permanent loss of every patient-player guarantee.

Repeated clean audits must eventually leave any fixed such rectangle, provided they occur while the relationship remains in it. With compliance odds \(L=\lambda/(1-\lambda)\) and capacity odds \(O\), the updates are \(L\mapsto L/(1-d)\) and \(O\mapsto Oq/\ah\). The first index at which either \(L(1-d)^{-n}\geq\Lambda/(1-\Lambda)\) or \(O(q/\ah)^n\leq O_I\) is finite. These formulas apply only through the first departure.

\subsection{Quantification and the patience limit}

The script \texttt{replication\_\allowbreak checks.py} in Online Resource 2 calculates thresholds using logarithms of odds, avoiding numerical rounding of a posterior to one. It verifies the firm upper bound using exact rational arithmetic, solves the auxiliary problem by recursion over successive clean-audit beliefs, and compares that solution with a separate Bellman iteration that optimizes over both effort choices. It also compares finite public-history trees with arbitrary admissible terminal values against the auxiliary solution. These checks supplement the proof; they neither solve the full equilibrium nor estimate its entry probability.

The calibration in the main text fixes \(\lambda_0=0.05\), \(\mu_0=0.5\), and \(\Lambda=0.1\). At \(\delta=0.99\), the sufficient capacity odds have base-ten logarithm approximately \(871.82\). Reporting the corresponding capacity posterior to ordinary decimal precision would misleadingly give one, although the theoretical cutoff is strictly below one. At \(\delta=0.999\), the construction gives \(M=7{,}748\) and \(K^P=140{,}888\). These magnitudes limit the quantitative interpretation of the all-equilibrium bound.

Finally,
\[
\lim_{\delta\uparrow1}a_\delta^\circ
=\frac{\al c}{\ell+\al c}<1,
\qquad
\lim_{\delta\uparrow1}\widehat\pi_\delta=q.
\]
Thus the admissible compliance-belief range need not vanish with patience, while the capacity belief needed to force a sufficiently long period of violation becomes increasingly pessimistic. The theorem concerns each discounted game separately. It does not make a statement about an undiscounted objective, an almost-sure entry event, or a uniform entry time as discounting vanishes.

\section{Capacity comparisons and welfare}\label{app-app:design}

This section supplies the calculations behind the capacity comparisons. The first calculation holds an effort rule fixed and asks how the capacity posterior responds to audit incidence. The next compares the sufficient silence bounds, whose thresholds can themselves change with the technology. We then derive losses over a common finite horizon on which the incentive conditions justify behavior. The calculation stopped at the first audit is presented separately because it uses a different stopping event.

\subsection{Posterior kernels and resource accounting}

For a persistent propensity with distribution \(G\), mean \(p\), and variance \(\nu\), the posterior kernels are
\begin{align}
dG_A(\theta)&=\frac{\theta}{p}\,dG(\theta),\label{app-eq:kernel-GA}\\
dG_{\No}(\theta)&=\frac{1-\theta}{1-p}\,dG(\theta).\label{app-eq:kernel-GN}
\end{align}
The posterior means are \((p^2+\nu)/p\) and \((p-p^2-\nu)/(1-p)\). The variance of the posterior mean across the two audit-incidence signals is \(\nu^2/[p(1-p)]\). These statements concern the same persistent propensity at both dates. If future effort changes, they need not describe the next realized audit rate. A clean audit gives the same capacity posterior only when its no-finding factor is common across capacity states.

For the binary distribution, \(\nu=\mu(1-\mu)(\ah-q)^2\). A fixed-mean change \(q'=q+\varepsilon\), \(\ah'=\ah-\mu\varepsilon/(1-\mu)\) reduces the spread by \(\varepsilon/(1-\mu)\). Thus
\[
p_{\No}'-p_{\No}=\frac{\nu-\nu'}{1-p}>0,
\qquad p_A'-p_A=-\frac{\nu-\nu'}p<0.
\]
If the two capacity states have construction costs \(k_C(q)\) and \(k_F(\ah)\), the expected cost derivative along this comparison is
\[
\frac{d}{d\varepsilon}\{
\mu k_C(q+\varepsilon)+(1-\mu)k_F(\ah-\mu\varepsilon/(1-\mu))\}
\bigg|_{\varepsilon=0}
=\mu\{k_C'(q)-k_F'(\ah)\}.
\]
It is zero under a common linear cost and nonpositive under a common differentiable convex cost. The convex-cost conclusion also follows directly from convex order without differentiability. Neither calculation specifies which capacity modifications are institutionally implementable.

\subsection{The specified high-effort path}

At fixed initial mean \(p\) and belief \(\mu\), put \(q(s)=p-(1-\mu)s\), \(\ah(s)=p+\mu s\). Following \(k\) high-effort periods with no audits,
\[
\frac{\mu_k(s)}{1-\mu_k(s)}
=\frac{\mu}{1-\mu}
\left(\frac{1-q(s)}{1-\ah(s)}\right)^k.
\]
For \(k\geq1\), \(\mu_k(s)>\mu\), and
\[
\frac{d}{ds}\log\frac{1-q(s)}{1-\ah(s)}
=\frac{1-\mu}{1-p+(1-\mu)s}+
\frac{\mu}{1-p-\mu s}>0.
\]
Consequently \(\mu_k'(s)>0\), and the maximum feasible next audit rate satisfies
\[
p_k^H(s)=p+s[\mu-\mu_k(s)],\qquad
\frac{d}{ds}p_k^H(s)=\mu-\mu_k(s)-s\mu_k'(s)<0.
\]
The comparison orders crossings of any common threshold. If high effort is also optimal at the successor, it orders actual audit rates along this specified history. It does not determine equilibrium effort before a sufficient region is reached.

\subsection{Why the two entry bounds must be distinguished}

For the original theorem, \(\pi_\delta\) and \(\beta_\delta\) depend on the payoffs, common low-effort rate, and discounting, but not on \(q,\ah\). If \(p>\pi_\delta\), the capacity target is
\[
\mu^\dagger(s)=\mu+\frac{p-\pi_\delta}{s}.
\]
Compression raises this target and lowers the minimum no-audit odds multiplier \(F(s)=[1-q(s)]/(1-\al)\). It therefore weakly lengthens the particular sufficient bound \(K^U\).

For the patient-player construction, \(\widehat\pi_\delta\) varies with \(q\), while the effort margin and hence the admissible comparison horizon can vary with \(\ah\). Even holding the horizon fixed does not rescue the old ordering. In the reversal example below, the two technologies \((q,\ah)=(0.02,0.66)\) and \((0.04,0.60)\) both have mean \(0.18\) at \(\mu=0.75\). The common low-effort rate is \(0.2\), \(\delta=0.9\), and \(\Lambda=0.1\). Their firm thresholds are \(0.09222222\) and \(0.10507937\); their auxiliary effort margins are \(1.31838441\) and \(1.42962309\). The smallest admissible horizon is \(M=28\) in both cases. The base-ten logarithms of \(O_I\) are \(44.93241180\) and \(34.98773877\), yielding \(K^P=505\) and \(436\). This is an explicit reversal for sufficient bounds, not for equilibrium outcomes.

For completeness, a single no-audit observation on a high-effort path crosses a common threshold \(\tau<p\) precisely when
\[
\nu>(p-\tau)(1-p).
\]
With \(\tau=\pi_\delta\) and \(\lambda<1-\beta_\delta\), this is a sufficient route into the original region. Reversing the inequality removes that route; it does not rule out violation or entry by other routes.

\subsection{Numerical comparisons of sufficient bounds}\label{app-app:bound-examples}

\begin{example}\label{app-ex:bound-comparison}
For the original bound, let \((g,f,c,r,\ell,d,\delta)=(1,4,0.1,1,1,0.9,0.1)\), \(\al=0.30\), and \((\lambda,\mu)=(0.50,0.75)\). Compare \((q,\ah)=(0.10,0.70)\) with \((q',\ah')=(0.20,0.40)\). Both maximum-effort means are \(0.25\), and the original thresholds are \(\pi_\delta=0.2173913\) and \(\beta_\delta=0.13580\). Following one high-effort no-audit period, the next audit ceilings are \(0.16\) and \(0.24\). Along the specified high-effort paths, \(\pi_\delta\) is crossed after one and five missed audits, respectively. The bounds that permit arbitrary pre-entry effort are \(K^U=2\) and \(K^U=10\). The specified high-effort paths have not been constructed as equilibria.

For the patient-player bound, instead use \((g,f,c,r,\ell,d,\delta)=(1,4,0.3,1,1,0.9,0.9)\), \(\al=0.2\), initial beliefs \((0.05,0.75)\), and the common cap \(\Lambda=0.1\). Compare \((q,\ah)=(0.02,0.66)\) with \((q',\ah')=(0.04,0.60)\). Both maximum-effort means are \(0.18\). The smallest comparison horizon in \eqref{eq:patient-M} is \(M=28\) for both technologies, but their bounds are \(K^P=505\) and \(K^P=436\). Thus compression shortens this particular bound. This reversal concerns the bounds, and establishes no reversal of actual equilibrium entry times.
\end{example}

\subsection{A common finite horizon with endogenous termination}

Suppose the chosen sufficient rectangle is \(\lambda<\Lambda\), \(O>\overline O\). To justify the first \(n\) decision periods, it suffices to check the all-clean path only through depth \(n-1\):
\[
\frac{\lambda}{1-\lambda}(1-d)^{-(n-1)}<\frac{\Lambda}{1-\Lambda},
\qquad O(q/\ah)^{n-1}>\overline O.
\]
The exponent is \(n-1\), since the action at the initial history is the first decision period. Induction proves the conditions at every active history of these depths. Beliefs after the final period may leave the region; no further behavior is used in the common-horizon calculation.

Conditional on a strategic firm and audit propensity \(\theta\), a no-audit period causes loss \(\mathcal H\), a clean audit causes loss \(\mathcal H+C_s\), and detection causes loss \(C_s+K_0\) and ends the relationship. Therefore expected current loss is \(\mathcal H+B\theta\), where \(B=C_s+d(K_0-\mathcal H)\), and conditional survival is \(1-d\theta\). Conditional on a committed firm, current loss is \(C_s\theta\) and survival is certain. This gives
\[
J_n^{\mathrm{pop}}(\theta)
=\lambda C_s\theta\sum_{t=0}^{n-1}\delta^t
+(1-\lambda)\sum_{t=0}^{n-1}\delta^t(1-d\theta)^t(\mathcal H+B\theta).
\]
The type weights in this ex-ante expression are the initial probabilities \(\lambda\) and \(1-\lambda\). Each type-conditional term already includes the probability of reaching its date without detection. Replacing the initial compliance weight by a survivor posterior would therefore count selection a second time. The distinction is the same one that appears in the population-share calculations below.

For \(t\geq1\), the curvature of the strategic term is
\[
\delta^t td(1-d\theta)^{t-2}
\{(t-1)d(\mathcal H+B\theta)-2B(1-d\theta)\}.
\]
Under \(B<0\), it is strictly positive; \(\mathcal H+B\theta>0\) follows from nonnegative real costs and \(d\theta<1\). At \(t=1\) the expression is simply \(-2\delta dB\). Hence strict convexity already holds for \(n=2\). For a nontrivial mean-preserving contraction, strict convexity gives a strict reduction in expected loss. With \(B=0\), the two-period kernel is affine and the comparison is an equality, while strict convexity holds from \(n=3\) onward. With \(B>0\), the two-period kernel is strictly concave, reversing the comparison.

For \(n=2\), all constant and linear terms cancel at a common mean, leaving
\[
\E_G[J_2^{\mathrm{pop}}]-\E_{G'}[J_2^{\mathrm{pop}}]
=(1-\lambda)\delta d(-B)(\nu-\nu').
\]
In the main-text example, the variances are \(0.007056\) and \(0.001692734694\), and the expected two-period losses are \(12.96459984\) and \(12.94357369\). The gain is \(0.02102614531\). The same discount factor, \(\delta=0.5\), is used for incentives and welfare. Both technologies satisfy the two-period guard, including the clean-audit branch.

\subsection{The alternative first-audit stopping calculation}

When a system starts in either sufficient region, all no-audit successors preserve the required behavior. A silence spell ending at the first audit, inclusive, is therefore another interval on which behavior is known. Its length differs across technologies, so it answers a different welfare question from the common-horizon comparison.

Conditional on a strategic firm,
\[
J^{\mathrm{spell}}_V(\theta)
=\frac{\mathcal H+B\theta}{1-\delta(1-\theta)}.
\]
Conditional on a committed firm, the corresponding loss is \(C_s\theta/[1-\delta(1-\theta)]\). Averaging over both types yields
\[
J^{\mathrm{spell}}_{\mathrm{pop}}(\theta)
=\frac{(1-\lambda)\mathcal H+B_{\mathrm{pop}}\theta}{1-\delta(1-\theta)},
\quad B_{\mathrm{pop}}=C_s+d(1-\lambda)(K_0-\mathcal H).
\]
For a numerator \(A+B_0\theta\), the first and second derivatives of this ratio are
\begin{align}
J'(\theta)&=\frac{B_0(1-\delta)-\delta A}{[1-\delta+\delta\theta]^2},\label{app-eq:Jprime}\\
J''(\theta)&=-\frac{2\delta[B_0(1-\delta)-\delta A]}{[1-\delta+\delta\theta]^3}.\label{app-eq:Jsecond}
\end{align}
Thus a strict capacity contraction lowers the stopped expected loss if \(B_0(1-\delta)<\delta A\). This condition must be checked using \(A=\mathcal H,B_0=B\) for strategic firms, or \(A=(1-\lambda)\mathcal H,B_0=B_{\mathrm{pop}}\) for the whole population. The conditional result cannot simply be relabeled as a population result.

The admissible parameters of the main-text welfare example satisfy both conditions. The expected population losses through the first audit are \(16.17751872\) and \(16.08781023\). These values include the first audit's loss, whether it is clean or detects a violation, and omit all later losses. No full-lifetime welfare ranking follows.

\section{Selection and the information retained by the observer}\label{app-app:survivors}

\subsection{One-period pooling}

Start at a common public history with independent types. Let \(H=1\) denote committed compliance and \(P=1\) constrained capacity, with initial probabilities \(\lambda\) and \(\mu\). Write \(a=(1-\lambda)\sigma\), \(p=\mu q+(1-\mu)x\), and \(D=1-dap\). The four type-pair survival probabilities are
\[
\begin{array}{c|cc}
 &P=1&P=0\\ \hline
H=1&1&1\\
H=0&1-d\sigma q&1-d\sigma x
\end{array}
\]
so the survivor marginals are \(\lambda/D\) and \(\mu(1-daq)/D\). The joint posterior of committed compliance and constrained capacity is \(\lambda\mu/D\). Consequently
\[
\Cov(H,P\mid S)
=\frac{\lambda\mu}{D}
-\frac{\lambda\mu(1-daq)}{D^2}
=\frac{\lambda\mu da(q-p)}{D^2}.
\]
When \(a>0\) and \(q<p\), survival favors both committed firms and constrained inspectors, and the covariance is negative. If \(a=0\), there is no detection risk under any pairing, so every relationship survives and neither type probability changes. This equality case shows why a difference in audit rates alone is insufficient for strict selection.

The covariance is calculated after combining the two nonterminal outcomes. If the observer instead learns which of \(\No\) and \(\Ac\) occurred, Lemma~\ref{app-lem:product-filter} gives independence at that public successor. Averaging the two conditional distributions changes the conditioning event and need not preserve independence. Thus the pooled-survival result and the full-history filtering result describe different information sets.

\subsection{Duration within a justified equilibrium horizon}

Suppose the finite-window guard holds for \(n\) periods. Then \(\sigma=1\) and \(x=\ah\) at every active decision history during that horizon in every equilibrium. For \(t\leq n\), define \(A_q=1-dq\), \(A_{\ah}=1-dh\). The four unnormalized survivor weights are
\[
w_{CP}=\lambda\mu,\quad w_{CF}=\lambda(1-\mu),\quad
w_{VP}=(1-\lambda)\mu A_q^t,\quad
w_{VF}=(1-\lambda)(1-\mu)A_{\ah}^t.
\]
Their sum is
\[
Z_t=\lambda+(1-\lambda)\{\mu A_q^t+(1-\mu)A_{\ah}^t\}.
\]
It follows that
\[
\lambda_t^S=\frac{\lambda}{Z_t},\qquad
\mu_t^S=\frac{\mu\{\lambda+(1-\lambda)A_q^t\}}{Z_t}.
\]
The covariance and odds ratio have the forms
\[
\Cov(H,P\mid S^t)
=\frac{\lambda\mu(1-\lambda)(1-\mu)(A_{\ah}^t-A_q^t)}{Z_t^2}<0,
\]
\[
\operatorname{OR}_t
=\frac{w_{CP}w_{VF}}{w_{CF}w_{VP}}
=\left(\frac{A_{\ah}}{A_q}\right)^t<1\qquad(t\geq1).
\]
For \(t=1\), these expressions agree with the one-period calculation after setting \(\sigma=1\).

Among strategic survivors the constrained odds are
\[
\frac{m_t}{1-m_t}=\frac{\mu}{1-\mu}\left(\frac{A_q}{A_{\ah}}\right)^t,
\]
which increase strictly. The harmful pair's population share is instead \(b_t=(1-\lambda)\mu A_q^t/Z_t\). Its change satisfies
\[
\frac{b_{t+1}}{b_t}=\frac{A_qZ_t}{Z_{t+1}}.
\]
Direct expansion gives
\[
A_qZ_t-Z_{t+1}
=d\{(1-\lambda)(1-\mu)(\ah-q)A_{\ah}^t-\lambda q\}.
\]
The expression in braces determines whether the harmful pair gains or loses population share. Its first term comes from the faster exit of strategic firms paired with functioning capacity; its second reflects the survival of committed firms relative to the positive exit risk of the harmful pair. The first term decreases strictly with duration. The share can therefore change from rising to falling at most once within the horizon over which the behavior is justified.

Equivalently, if \(\chi_t=1-Z_{t+1}/Z_t\) is the next detection hazard among survivors, then \(b_t\) rises precisely when \(\chi_t>dq\). Within the constant-behavior horizon, selection weights the persistent pair-specific hazards \(0,dq,dh\) by their one-period survival probabilities. Therefore
\[
\chi_{t+1}=\chi_t-\frac{\Var(\text{pair-specific hazard}\mid S^t)}{1-\chi_t}<\chi_t,
\]
whenever both hazards being compared lie within the justified decision horizon. This is a heterogeneity-and-selection identity. It is not separate evidence that effort has fallen.

The original technology in the main-text welfare illustration gives
\[
(b_0,b_1,b_2)=(0.97020,0.98075334,0.98442580),
\]
\[
(m_0,m_1,m_2)=(0.98,0.99177046,0.99663750).
\]
With only the initial compliance belief changed to \(\lambda=0.2\), initial membership in the original sufficient region still holds. The one-period harmful share becomes \(b_1=0.77601810<b_0=0.784\). This second calculation uses only one period of justified behavior; it does not claim that the two-period guard survives the change in \(\lambda\).

\subsection{Full silence records and limits of extrapolation}

After a history in a sufficient region, further no audits keep the relationship in it for every finite silence length. Conditional on \(\No^t\), the four likelihoods depend only on inspector type: they are \((1-q)^t\) for constrained capacity and \((1-\ah)^t\) for functioning capacity. Hence the compliance posterior remains \(\lambda\), types remain independent, and the harmful-pair share is
\[
(1-\lambda)\frac{\mu(1-q)^t}{\mu(1-q)^t+(1-\mu)(1-\ah)^t}.
\]
This share increases to \(1-\lambda\) as a sequence of conditional probabilities. It does not assert positive probability of an infinite silence run; every audit propensity is positive.

If violation and high effort are imposed forever, the duration formulas define a fixed-behavior benchmark for every \(t\). In that benchmark,
\[
Z_t\longrightarrow\lambda,\qquad
m_t\longrightarrow1,\qquad
b_t\longrightarrow0,\qquad
\lambda_t^S\longrightarrow1,\qquad
\mu_t^S\longrightarrow\mu.
\]
In particular, the constrained-inspector marginal is not globally increasing: it initially exceeds \(\mu\), but eventually returns to \(\mu\) as committed firms dominate the remaining sample. These benchmark limits cannot be read as equilibrium limits when clean audits leave the dominance region. They show why a shrinking survivor-sample odds ratio never by itself establishes a growing harmful share in the whole population.

\section{Numerical verification and robustness}\label{app-app:verification}

We report numerical values for the terminal construction and the simple global bound, followed by the capacity comparison. The accompanying scripts check the filtering formulas, continuation bounds, finite-history comparisons, welfare calculations, and survivor identities. These calculations provide checks on the analytical results. They do not determine equilibrium play outside the sufficient regions or estimate the probability of entry in an unrestricted equilibrium.

\subsection{Two-period example}

For
\[
(g,f,c,r,\ell,\delta,d)
=(1,4,0.3,1,1,0.9,0.98)
\]
and
\[
(q,\al,\ah,\lambda,\mu)
=(0.02,0.10,0.98,0.05,0.72),
\]
the exact formulas give
\[
p^*=0.2040816327,\qquad
a^*=0.1530612245,
\]
\[
\widehat a=0.4584434855,\qquad
\rho_0=0.6561754572,\qquad
\sigma_0=0.4825720900.
\]
After a clean audit,
\[
\rho_{\Ac}=0.1341551128,\qquad
\sigma_{\Ac}=0.1683451735.
\]
After no audit,
\[
\mu_{\No}=0.8865230769,\qquad
p_{\No}=0.1289378462.
\]
Substitution into both first-period deviation differences gives zero up to machine precision.

\subsection{Global example}

For
\[
(g,f,c,r,\ell,\delta,d)
=(1,4,0.3,1,1,0.5,0.9)
\]
and
\[
(q,\al,\ah,\lambda_0,\mu_0)
=(0.02,0.20,0.80,0.10,0.50),
\]
the calculations are
\[
\kappa_L=0.7000000000,\qquad
R=0.7000000000,
\]
\[
\pi_\delta=0.1851851852,\qquad
\beta_\delta=0.7500000000,
\]
\[
\mu^\dagger=0.7882241216,\qquad
F=1.225,\qquad K=7.
\]
The lower-bound posterior at entry is \(0.8054302519\), giving maximum audit rate \(0.1717644035\). After another no-audit observation, the corresponding values are \(0.9530158450\) and \(0.0566476409\).

\subsection{Design example and code record}

With \(\mu=0.75\), the two capacity distributions
\[
(q,\ah)=(0.10,0.70)
\quad\text{and}\quad
(q',\ah')=(0.20,0.40)
\]
both have mean \(0.25\). Their variances are \(0.0675\) and \(0.0075\). Post-silence means are \(0.16\) and \(0.24\); post-audit means are \(0.52\) and \(0.28\). The critical variance for \(\pi_\delta=0.2173913043\) is \(0.0244565217\). On the maximum-effort path, the threshold is crossed after one no-audit observation under the dispersed technology and after five under the compressed technology: the latter has
\[
p_4'=0.2190812721,\qquad p_5'=0.2146606335.
\]
Taking \(\al=0.30\), the uniform every-equilibrium entry bounds are
\[
K^U(s)=2,\qquad K^U(s')=10.
\]

Appendix~\ref{app-app:design} reports the common-horizon welfare example. The script \texttt{replication\_\allowbreak checks.py} in Online Resource 2 uses exact rational arithmetic for the capacity paths, finite-window boundary cases, welfare recursions, and survivor formulas. It also reproduces the patient-player bound comparison using log odds. The verification records specify which identities and examples were checked.

\subsection{What is and is not robust}

Since every inequality displayed in the two-period construction is strict, the exact equilibrium is continuous in \((d,q,\al,\ah)\) and therefore persists under small perturbations of implementation and detection noise.

The simple global theorem permits arbitrary equilibrium play and dependence on private action histories. Its likelihood-ratio argument nevertheless relies on capacity separation, and its continuation bounds are conservative. Reversing the capacity ordering removes the uniform learning inequality; it does not by itself establish a particular equilibrium outcome under overlapping technologies. Similarly, failure of a sufficient dominance inequality leaves the action at that history unresolved. For sufficiently patient players, \(\pi_\delta\) approaches zero and \(\beta_\delta\) exceeds one, so the simple sufficient region is unavailable. Appendix~\ref{app-app:patient} supplies the different argument for every fixed discount factor below one, with a bound that is not uniform in the patience limit.

The capacity results require a specified effort rule for posterior propensity to represent the next actual audit rate. The original sufficient-bound ordering does not extend automatically to the patient-player construction; the explicit reversal illustrates the distinction. Welfare is compared over a common finite horizon only when the all-clean guard justifies behavior at every active history under both technologies. The alternative silence-spell calculation uses a different stopping event and is not a lifetime comparison.

The duration formulas likewise require behavior to be justified over every pooled public history. Outside that finite window, the constant-behavior formulas are only a benchmark. Negative type association among survivors does not imply a rising population share of the strategic/constrained pair. Full audit histories retain the separate product-form posterior.

\end{document}